\documentclass[fleqn,usenatbib,useAMS]{mnras}

\usepackage[T1]{fontenc}
\usepackage{ae,aecompl}
\usepackage[normalem]{ulem}

\usepackage{graphicx}	
\usepackage{amsmath}	
\usepackage{amssymb}	
\usepackage{geometry}
\usepackage{float}
\usepackage{hyperref}
\usepackage{breqn}
\usepackage{mathtools}
\usepackage{multirow}
\usepackage{cite}

\newcommand{\kms}{\,\rm{km}\,\rm{s}$^{-1}$}

\newcommand{\shark}{\textsc{Shark}} 
\newcommand{\lcdm}{$\Lambda$CDM}
\newcommand{\surfs}{\textsc{surfs}}
\newcommand{\M}{$\rm{M}_{\odot}$}
\newcommand{\vmax}{$V_{\rm{max}}$}
\newcommand{\wfifty}{$W_{50}$}
\newcommand{\wtwenty}{$W_{20}$}
\newcommand{\velociraptor}{\textsc{velociraptor}\ }
\newcommand{\hi}{\textsc{Hi}}
\newcommand{\hmol}{H$_{2}$}

\title[The \hi\ velocity function]{The \textmd{\hi}\ Velocity Function: a test of cosmology or baryon physics?}

\author[G. Chauhan et al.]{Garima Chauhan,$^{1,2}$\thanks{E-mail: garima.chauhan@icrar.org}
Claudia del P. Lagos,$^{1,2}$
Danail Obreschkow,$^{1,2}$
Chris Power,$^{1,2}$
\newauthor{Kyle Oman,$^{3}$
Pascal J. Elahi,$^{1,2}$}
\\
$^{1}$International Centre for Radio Astronomy Research (ICRAR), 7 Fairway, Crawley, WA 6009, Australia.\\
$^{2}$ARC Centre of Excellence for All Sky Astrophysics in 3 Dimensions (ASTRO 3D), Australia.\\
$^{3}$Kapteyn Institute,Landleven 12, 9747 AD Groningen, Netherlands.
}

\date{Accepted XXX. Received YYY; in original form ZZZ}

\pubyear{2019}

\begin{document}
\label{firstpage}
\pagerange{\pageref{firstpage}--\pageref{lastpage}}
\maketitle

\begin{abstract}
Accurately predicting the shape of the \hi\ velocity function of galaxies is regarded widely as a fundamental test of any viable dark matter model. Straightforward analyses of cosmological $N$-body simulations imply that the \lcdm\ model predicts an overabundance of low circular velocity galaxies when compared to observed \textmd{\hi} velocity functions. More nuanced analyses that account for the relationship between galaxies and their host haloes suggest that how we model the influence of baryonic processes has a significant impact on \hi\ velocity function predictions. We explore this in detail by modelling \hi\ emission lines of galaxies in the \shark\ semi-analytic galaxy formation model, built on the \surfs\ suite of \lcdm\ $N$-body simulations. We create a simulated ALFALFA survey, in which we apply the survey selection function and account for effects such as beam confusion, and compare simulated and observed \hi\ velocity width distributions, finding differences of $\lesssim 50$\%, orders of magnitude smaller than the discrepancies reported in the past. This is a direct consequence of our careful treatment of survey selection effects and, importantly, how we model the relationship between galaxy and halo circular velocity - the \hi\ mass-maximum circular velocity relation of galaxies is characterised by a large scatter. These biases are complex enough that building a velocity function from the observed \hi\ line widths cannot be done reliably. 
\end{abstract}






\section{Introduction}
\label{sec:Introduction}

The $\Lambda$ Cold Dark Matter (hereafter \lcdm) model is well established as the Standard Cosmological Model, naturally predicting  the structure of the Universe on intermediate-to-large scales and explaining a swathe of observational data, from the formation and evolution of large scale structure, to the state of the Early Universe, to the cosmic abundance of different types of matter (e.g. \citealt{Bull_2016}). 

Despite its numerous successes, however, the \lcdm\ model faces a number of challenges on small scales. Cold dark matter (hereafter CDM) haloes form cuspy profiles \citep[i.e. the dark matter density rises steeply at small radii][] {Navarro1995-Xray}, whereas observational inferences suggest that low mass dark matter (hereafter DM) dominated galaxies have constant-density DM cores \citep{Duffy2010, Oman2015, Dutton2018NIHAOFunction}, leading to the so-called ``cusp-core" problem. CDM haloes are also predicted to host thousands of subhaloes, which has led to the conclusion that the Milky Way suffers from a ``missing satellites'' problem because it should host many more satellite galaxies than the $\sim 50$ that are observed \citep{Bullock2017Small-ScaleParadigm}. While the inefficiency of galaxy formation in low-mass haloes - because of feedback processes such as  e.g. cosmological reionization, supernovae, etc... - may lead to many subhaloes to be free of baryons and dark, the ``too big to fail'' problem  \citep{Boylan-Kolchin-2011} suggests that the central density of CDM subhaloes are too high; in dissipationless \lcdm\ simulations of Milky Way mass haloes, the most massive subhaloes, which are large enough to host galaxy formation and so ``too big to fail'', have typical circular velocities ~1.5 times higher ($\sim 30$~\kms) than that observed at the half-light radii of the Milky Way satellite. This indicates that there are problems with both the predicted abundances and internal structures of CDM subhaloes \citep{Dutton2016NIHAOHaloes}.

Interestingly, with the emergence of observational surveys sensitive enough to detect statistical samples of faint galaxies in the nearby Universe, it has become clear that there is a consistent deficit in the observed abundance of low mass galaxies when compared to predictions from the \lcdm\ model \citep[e.g.][]{Tollerud-2008,Hargis-William-Peter_2014}. This suggests that the ``missing satellite'' problem is more generically a ``missing dwarf galaxy'' problem. This is most evident in measurements of the velocity function (VF) - the abundance of galaxies as a function of their circular velocity. The observed VF is assumed to be equivalent to the VF of DM subhaloes \citep{Gonzalez2000}, and so its measurement should provide a potentially powerful test of the Standard Cosmological Model. 

The utility of the VF as a test of DM is already evident in the results of the \hi\ VF measured by ALFALFA (The Arecibo Legacy Fast ALFA: \citealp{Giovanelli2005}); focusing on galaxies with rotational velocities of $\sim 25$~\kms, the ALFALFA VF found approximately an order of magnitude fewer galaxies than expected from cosmological CDM simulations \citep{Klypin2014,Brooks2017}. \citet{Trujillo-Gomez2018} attempted to correct the measured \hi\ velocities by including the effects of pressure support and derive a steeper VF, though still shallower than the \lcdm\ prediction.
This has prompted interest in Warm Dark Matter (hereafter WDM) models, which predict significantly less substructure within haloes \citep{Maccio-WDM-2012,Zavala2009}. The linear matter power spectrum in WDM cosmologies is characterised by a steep cutoff at dwarf galaxy scales, which results in the suppression of low-mass structure formation and a reduction in the number of dwarf galaxies such that the VF predicted by the WDM model is more consistent with observations \citep{Schneider-2012}. While the WDM model has the potential to provide a better description of the observed VF, there is a tension between the range of WDM particle masses required \citep[$<$ 1.5 keV; cf.][]{Schneider2016} and independent observational constraints from the Lyman-$\alpha$ forest at high redshifts, which rule out such low WDM particle masses \citep{Klypin2014}. 

An alternative solution that has been recently discussed to alleviate the discrepancy between the observed and predicted VF is the effect of baryonic physics. \citet{Brooks2017} and \citet{Maccio2016} used cosmological zoom-in hydrodynamical simulations of a small number of galaxies (typically ranging from $30$ to $100$) to produce \hi\ emission lines for their galaxies. They measured \wfifty\ (width of the \hi\ emission line at $50$\% of the maximum peak flux), which is used as a proxy in observations to estimate the \hi\ velocity of the galaxy, and then compared them with the rotational velocity, $V_{\rm \textsc{dmo}}$, of the haloes from the dark matter only (DMO) simulations. They found that due to the effect of baryons, \wfifty\ and $V_{\rm \textsc{dmo}}$ are non-linearly correlated, in a way that \wfifty\ tends to underestimate $V_{\rm \textsc{dmo}}$ in low mass haloes, while the opposite happens at the high-mass end. They propose that a DM density profile that varies with stellar-to-halo mass ratio can be used to reconcile the differences with the observations. 
{\citet{Trujillo-Gomez2018}, however, showed that including the feedback-induced deviations from the \lcdm\ VF predicted by the hydrodynamical simulations above were insufficient to reproduce the observed VF.} 

Although the work of \citet{Brooks2017} and \citet{Maccio2016} present a compelling solution to the apparent missing dwarf galaxy problem, their sample is statistically limited. \citet{Obreschkow2013} approached this problem from a different perspective, with much better statistics (going into a million of simulated galaxies). They attempted to see how the selection biases of the surveys might contribute to this problem. Their solution was to make a mock-survey using DMO $N$-body simulations combined with semi-analytic models of galaxy formation, and then compare its results with the actual observations via producing a lightcone (see $\S$ \ref{sec:Lightcone}) with all the required selection effects. They did this for the HIPASS survey (\hi\ Parkes All-Sky Survey: \citealp{Meyer2004}), as their simulation was limited in resolution to moderate halo masses, and hence was more directly comparable to HIPASS. HIPASS is the first blind \hi\ survey in the Southern Hemisphere with a velocity range of $-1280$ to $12700$ \kms, identifying over $5317$ \hi\ sources in total (including both Northern and Southern Hemispheres). \citet{Obreschkow2013} found that the observed \hi\ linewidths were consistent with \lcdm\ at the resolution of the Millennium simulation \citep{Springel2005}, though they could not comment on haloes of lower mass, in which the largest discrepancies have been reported. 

The main limitations of the works above have been either statistics or limited resolution. Here, we approach this problem with the \surfs~suite \citep{Elahi_SURFS} of $N$-body simulations, which covers a very large dynamic range, from circular velocities of $20$~\kms\ to $>500$~\kms, and combine it with the state-of-the-art semi-analytic model \shark\ \citep{Lagos2018Shark:Formation}, which includes a sophisticated multi-phase interstellar medium modelling. We use these new simulations and model to build upon the work of \citet{Obreschkow2013}, and present a thorough comparison with the $100\%$-data release of ALFALFA \citep{Haynes2018TheCatalog}. We focus on the ALFALFA survey as it is a blind \hi\ survey and covers a greater cosmological volume with a better velocity and spatial resolution than other previous \hi\ surveys. We show that our simulated ALFALFA lightcone produces a \wfifty\ distribution in very good agreement with the observations, even down to the smallest galaxies detected by ALFALFA, and discuss the physics behind these results and their implications.

This paper is organised as follows. $\S$~\ref{sec:Galaxy Catalogue} describes the galaxy formation model used in this study and the construction of the mock ALFALFA survey. In $\S$~\ref{sec:Modelling}, the modelling of the \hi\ emission lines is described along with its application on the mock-sky built in the previous section. $\S$~\ref{sec:Comparisons}, we compare our results with ALFALFA observations and discuss our results in the context of previous work. $\S$~\ref{sec:conclusions} summarises our main results. In the Appendix \ref{sec:appendix} we compare our model for the \hi\ emission line of galaxies with the more complex \hi\ emission lines obtained from the cosmological hydrodynamical simulations  \textsc{apostle}  \citep{Oman2019}.


\section{The simulated galaxy catalogue}
\label{sec:Galaxy Catalogue}

\begin{figure*}
	\includegraphics[trim= 1cm 19mm 2cm 10mm,width=0.85\linewidth]{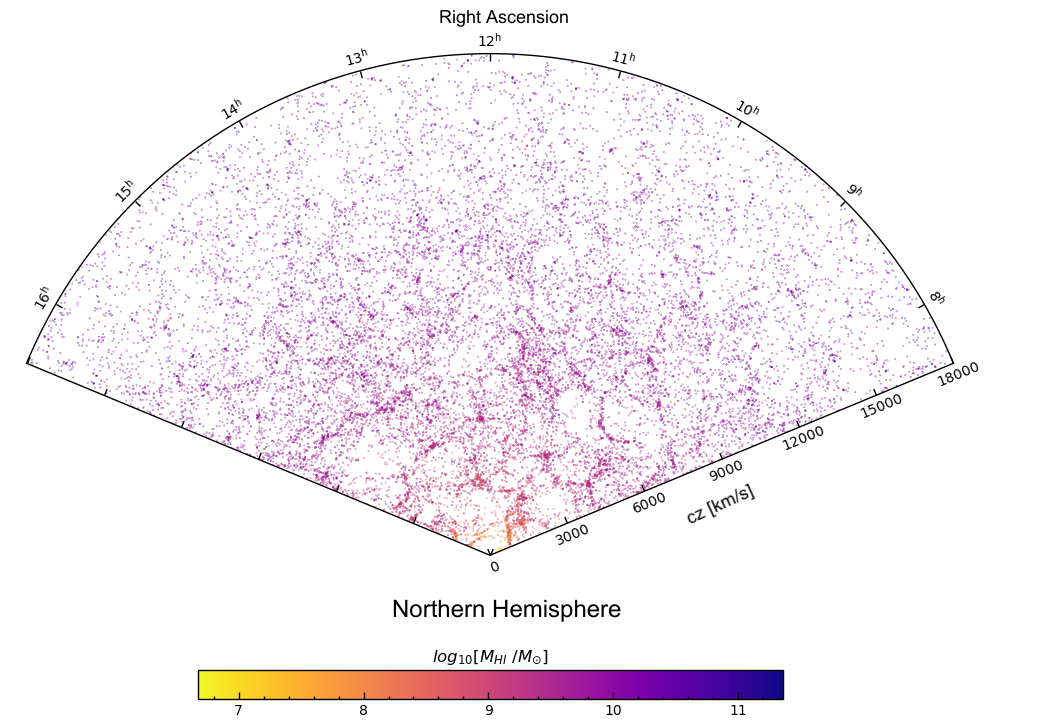}
	\includegraphics[trim= 1cm 5mm 2cm 10mm,width=0.89\linewidth]{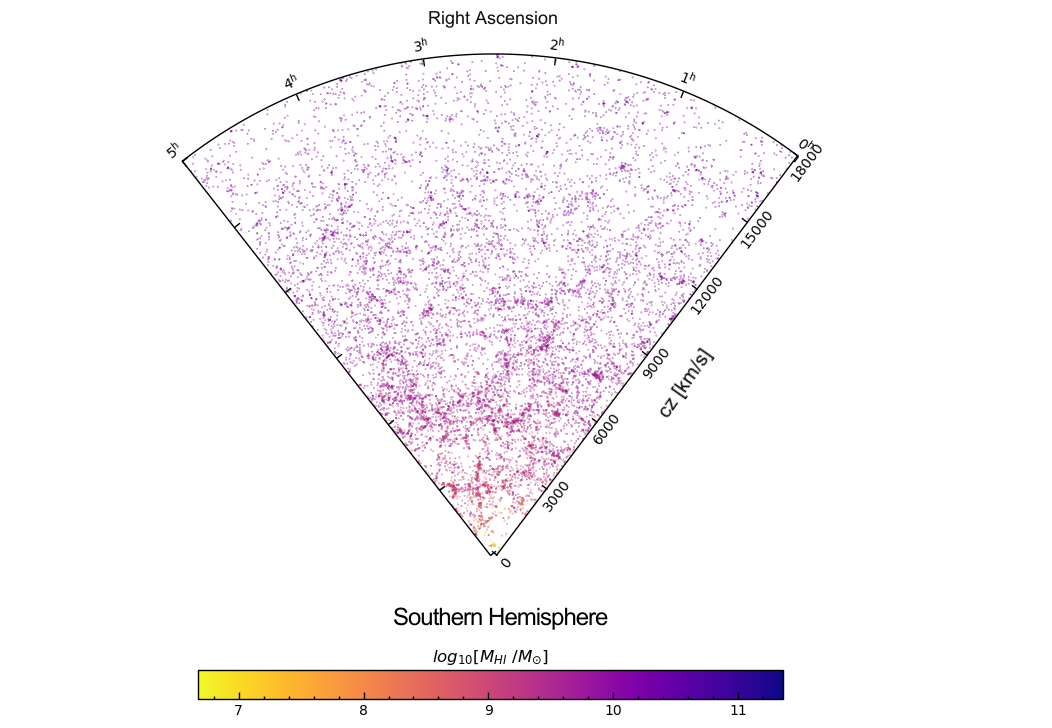}
	\caption{Mock sky of the ALFALFA survey, created with the outputs of \shark\ and processed with Stingray to create the observable sky. Symbols show individual galaxies and colours show their \hi\ mass, as labelled by the colour bar at the bottom. Low \hi\ mass galaxies are only detected in the very nearby universe.}
	\label{fig:SAM}
\end{figure*}



Our simulated galaxy catalogue is constructed using the \shark\ semi-analytic model \citep{Lagos2018Shark:Formation} that was run on the \surfs\ $N$-body simulations suite \citep{Elahi_SURFS}. Here, we describe briefly \shark\ and \surfs.

Hierarchical galaxy formation models, such as \shark, require three basic pieces of information about DM haloes : (i) the abundance of haloes of different masses; (ii) the formation history of each halo; and in some cases (iii) the internal structure of the halo including their radial density and their angular momentum \citep{Baugh2006}. These fundamental properties are now well established, thanks to the $N$-body simulations like \surfs\ (used in this study).

The \surfs\ suite consists of $N$-body simulations of differing volumes, from $40\,h^{-1}$~cMpc to $210\,h^{-1}$~cMpc on a side, and particle numbers, from $\sim\! 130$ million up to $\sim\!8.5$ billion particles, using the \lcdm~\textit{Planck} cosmology (Planck Collaboration XIII \citeyear{PlanckXIII}). The latter has a total matter, baryon and dark energy densities of $\Omega_{\rm m}=0.3121$, $\Omega_{\rm b}=0.0491$ and $\Omega_{\rm L}=0.6751$, and a dimensionless Hubble parameter of $h=0.67512$. The \surfs\ suite is able to resolve DM haloes down to $8.3\times 10^8\,h^{-1}$ \M. For this analysis, we use the L40N512 and L210N1536 runs, referred to as micro-\surfs\ and medi-\surfs\, respectively hereafter, whose properties are given in Table \ref{tab:sims}. Merger trees and halo catalogues were constructed using the phase-space finder {\sc velociraptor} \citep{Elahi2019-Velociraptor,Canas2018} and the halo merger tree code {\sc TreeFrog} \citep{Poulton_Treefrog2018,Elahi2019-TreeFrog}. 

\shark\ was introduced by \citet{Lagos2018Shark:Formation}, and is an open source, flexible and highly modular cosmological semi-analytic model of galaxy formation, which is hosted in GitHub\footnote{\href{https://github.com/ICRAR/shark}{\url{https://github.com/ICRAR/shark}}}. It models key physical processes that shape the formation and evolution of galaxies, including (i) the collapse and merging of DM haloes; (ii) the accretion of gas onto haloes, which is {governed by} the DM accretion rate; (iii) the shock heating and radiative cooling of gas inside DM haloes, leading to the formation of galactic discs via conservation of specific angular momentum of the cooling gas; (iv) the formation of a multi-phase interstellar medium and star formation in galaxy discs; (v) the suppression of gas cooling due to photo-ionisation; (vi) chemical enrichment of stars and gas; (vii) stellar feedback from the evolving stellar populations; (viii) the growth of supermassive black holes via gas accretion and merging with other black holes; (ix) heating by active galactic nuclei (AGN); (x) galaxy mergers driven by dynamical friction within common DM haloes which can trigger bursts of star formation (SF) and the formation and/or growth of spheroids; and (xi) the collapse of globally unstable discs that also lead to the bursts of SF and the formation and/or growth of bulges. \shark\ includes several different models for gas cooling, AGN feedback, stellar and photo-ionisation feedback, and star formation. The model also numerically evolves the exchange of mass, metals and angular momentum between the key gas reservoirs of haloes and galaxies: halo hot and cold gas, galaxy stellar and gaseous' disc and bulge (and within discs between the atomic and molecular gas), central black hole, and the ejected gas component (outside haloes).

\begin{table}
\makeatletter
 \def\@textbottom{\vskip \z@ \@plus 1pt}
 \let\@texttop\relax
\makeatother

        \setlength\tabcolsep{2pt}
        \centering\footnotesize
        \caption{\surfs\ simulation parameters of the runs being used in this paper. We refer to L40N512 and L210N1536 as micro-\surfs\ and medi-\surfs, respectively.}
        \begin{tabular}{@{\extracolsep{\fill}}l|cccc|p{0.45\textwidth}}
                \hline
                \hline
                Name & Box size & Number of & Particle Mass & Softening Length\\
                & $L_{\rm box} [\rm cMpc/h]$ & Particles $N_{\rm p}$ & $m_{\rm p}$ [\rm \M/h]  & $\epsilon [\rm ckpc/h]$ \\
                 \hline
                L40N512     & $40$  & $512^3$   & $4.13\times10^7$ & 2.6  \\
                L210N1536   & $210$ & $1536^3$  & $2.21\times10^8$ & 4.5  \\
        \end{tabular}
        \label{tab:sims}
        
\end{table}


Halo gas in \shark\ is assumed to be in two phases: cold, which is expected to cool within the duration of a halo's dynamical time; and hot, which remains at the virial temperature of the halo. Cold gas is assumed to settle onto the disc and follows an exponential profile of half-mass radius $r_{\rm gas, disc}$. In our model $r_{\rm gas, disc}$ can differ from the stellar half-mass radius as stars form only from the molecular hydrogen (\hmol) and not the total gas. Surface densities of \hi\ and \hmol\ are calculated using the pressure relation of \citet{Blitz2006}, described in detail in $\S$~\ref{sec:gasprofile}. 

Models and parameters used in this study are the defaults of \shark\ as described in \citet{Lagos2018Shark:Formation}, which were calibrated to reproduce the $z=0,\,1,\,2$ stellar mass functions (SMFs); the $z=0$ black hole-bulge mass relation; and the disc and bulge mass-size relations. In addition, the model reproduces well observational results that are independent of those used in calibration, including the total neutral, atomic and molecular hydrogen-stellar mass scaling relations at $z=0$; the cosmic star formation rate (SFR) density evolution up to $z\approx 4$; the cosmic density evolution of the atomic and molecular hydrogen at $z\lesssim 2$ or higher in the case of the latter; the mass-metallicity relations for the gas and stars; the contribution to the stellar mass by bulges and the SFR-stellar mass relation in the local Universe. \citet{Davies2018} show that \shark\ also reproduces the scatter around the main sequence of star formation in the SFR-stellar mass plane, while Martindale et al. (in preparation) show that \shark\ reproduces the \hi\ content of groups as a function of halo mass. Of particular importance for this study is \shark's success in recovering the observed gas abundances of galaxies.


\subsection{A mock ALFALFA sky}
\label{sec:Lightcone}

To ensure a fair comparison with available \hi\ surveys, we first estimate how predicted galaxy properties are likely to be influenced by the choice of selection criterion. Here, mock galaxy catalogues are a particularly powerful tool, and so we begin by constructing a ``mock ALFALFA" survey. We do this by generating a galaxy population with \shark\ and embed them within a cosmological volume by applying the survey's angular and radial selection functions (e.g. \citealp{Merson2012}). 

{We use the code \textsc{stingray}, which is an extended version of the lightcone of \citet{Obreschkow_2009_ligthcone}, to build our lightcones from the \shark\ outputs.}
{Rather than forming a single chain of replicated simulation boxes, \textsc{stingray} tiles boxes together to build a more complex 3D field along the line-of-sight of the observer.} Galaxies are drawn from simulation boxes which correspond to the closest look-back time, which ranges over the redshift range $z=0$ to $z=0.06$ (corresponding to the ALFALFA limit); in the \shark\ simulations, this corresponds to the last 7 snapshots. Properties of each galaxy in the lightcone are obtained from the closest available time-step, resulting in the formation of spherical shells of identical redshifts. {A possible issue would be the same galaxy appearing once in every box, but due to cosmic evolution might display different intrinsic properties.}
In order to avoid this problem, galaxy positions are randomised by applying a series of operations consisting of $90$~deg-rotations, inversions, and continuous translations.
We build the lightcones with all the galaxies in \shark\ that have a stellar or cold gas mass (atomic plus molecular) $\ge 10^6\,\rm M_{\odot}$. Any additional selection (in this case the one specific to ALFALFA) are applied later, directly to the lightcone galaxies. The end result of the whole process is that we get a mock-observable sky as shown in Figure~\ref{fig:SAM} which is as near to the real sky as possible and with minimum repetition of the large-scale structure. The two portions of the sky shown correspond to the north and south ALFALFA regions.

{\sc stingray} also computes an inclination for each galaxy with respect to the observer. The latter are constructed assuming galaxies to have an angular momentum vector of the same direction as of its subhalo angular momentum vector (as measured by \velociraptor), in the case of central galaxies and satellites galaxies type =1. For type=2 satellite galaxies we assume random orientations. Satellites type=1 correspond to those hosted by satellite subhalos that are identified by {\sc velociraptor}, while satellites type=2 correspond to those that were hosted by subhalos that have ceased to be identified by {\sc velociraptor}. The latter usually happens when subhalos become too low mass to be robustly identified (see \citealt{Poulton_Treefrog2018} for a detailed analysis of satellite subhalo orbits). The overall effect of inclinations is to reduce \wfifty.

A limitation of any observational survey is finite velocity and spatial resolution, which for a survey like ALFALFA can lead to 2 or more galaxies falling inside the same beam and then overlapping in frequency, more commonly known as {``beam confusion''}. To mimic the effect of {confusion} in our analysis, we merge simulated galaxies whose centroids are separated by less than a projected 3.8' (the full-width-half-max for the ALFALFA beam) and whose \hi\ lines overlap in frequency. In the case of galaxies being confused, the common \hi\ mass is taken as the sum of the individual \hi\ masses of the galaxies, and the \wfifty\ (the full-width at half of the peak flux of the line)
is measured for the combined line formed due to the overlapping \hi\ lines. {\citet{Obreschkow2013} found that ``confused" galaxies typically have high \hi-mass and \wfifty, with $M_{\rm HI} > 10^{10}$\M\ and $W_{50} > 300$\kms, albeit for the HIPASS survey, which has a larger beam than ALFALFA; we find fewer confused galaxies lying in this range in our sample. By including confusion, we reduce the total number of galaxies by $<1$\%, throughout the whole dynamical range of galaxies.} 

To ensure that we have the dynamical range in circular velocity in our sample of galaxies required to test the ``missing satellite problem'', we make two lightcones using the micro- and medi-\surfs; micro-\surfs\ gives us better mass resolution to probe down to dwarf galaxies, with $M_{\rm HI} \lesssim 10^9 M_{\odot}$, while medi-\surfs\ provides us with a much larger volume and better statistics at the high-mass end, $M_{\rm HI}\gtrsim 10^9 M_{\odot}$. Results for these lightcones are presented in \S~\ref{sec:Reproducing ALFALFA}.


\section{Modelling \textmd{\hi} emission lines in Galaxy Formation Models}
\label{sec:Modelling}

In this section, we describe the steps required to build an \hi\ emission line for each \shark\ galaxy. $\S$~\ref{sec:gasprofile} and $\S$~\ref{sec:velocity_profile} provide details of the surface density and velocity profile calculations, respectively. The way we combine them to create the \hi\ emission line is described in $\S$~\ref{sec:emissionline}.


\subsection{Gas mass and profile}
\label{sec:gasprofile}

For the calculation of the \hi\ surface density profile, we adopt the empirical model described in \citet{Blitz2004TheFormation,Blitz2006} (Equation~\ref{equation:BR06}). In their model, the ratio of molecular to atomic hydrogen gas surface density in galaxies is a function of hydro-static pressure in the mid-plane of the disc, with a power-law index close to $1$,  

\begin{equation}
     R_{\rm mol}(r) = \left[P_{\rm ext}(r)/P_{\star}\right]^{\alpha},
    \label{equation:BR06}
\end{equation}

\noindent where $R_{\rm mol}\equiv \Sigma_{\rm H_{2}}/ \Sigma_{\rm HI}$, with $\Sigma_{\rm H_{2}}$ and $\Sigma_{\rm HI}$ being the surface density of molecular and atomic hydrogen, respectively. The parameters $P_{\star}$ and $\alpha$ are measured in observations, and in \shark\ we adopt $P_{\star} = 34,673{\rm \,K cm^{-3}}$ and $\alpha =0.92$, which correspond to the best fit values in \citet{Blitz2006}.

\citet{Blitz2006} adopted the \citet{Elmegreen1989} estimate of $P_{\rm ext}$ for disc galaxies, which corresponds to the mid-plane pressure in an infinite, two-fluid disc with locally isothermal stellar and gas layers, 

\begin{equation}
P_{\rm ext}(r) = \frac{\pi}{2} G \Sigma_{\rm g} \left[\Sigma_{\rm g} + \left(\frac{\sigma_{\rm gas}}{\sigma_{\star}}\right)\Sigma_{\star}\right],
\label{equation:BR06 - pext}
\end{equation}

\noindent where $P_{\rm ext}(r)$ is the kinematic mid-plane pressure outside molecular clouds, and the input for Equation~\ref{equation:BR06}. $G$ is the gravitational constant, $\Sigma_g$ is the total gas surface density (atomic plus molecular), $\Sigma_{\star}$ is the stellar surface density, and $\sigma_{\rm gas}$ and $\sigma_{\star}$ are the gas and stellar vertical velocity dispersion, respectively.

The stellar and gas surface densities are assumed to follow exponential profiles with a half-gas and half-stellar mass radii of $r_{\rm gas, disc}$ and $r_{\star,\rm disc}$, respectively. We adopt $\sigma_{\rm gas}=10\,\rm km\,s^{-1}$ \citep{Leroy2008} and calculate $\sigma_{\star}=\sqrt{\pi\,G\,h_{\star}\,\Sigma_{\star}}$. Here, $h_{\star}$ is the stellar scale height, and we adopt the observed relation $h_{\star}=r_{\rm \star,disc}/7.3$ \citep{Kregel2002FlatteningGalaxies}, with $r_{\rm \star, disc}$ being the half-stellar mass radius. 

\begin{figure}
	\includegraphics[trim=0mm 15mm 0mm 0mm ,width=\linewidth]{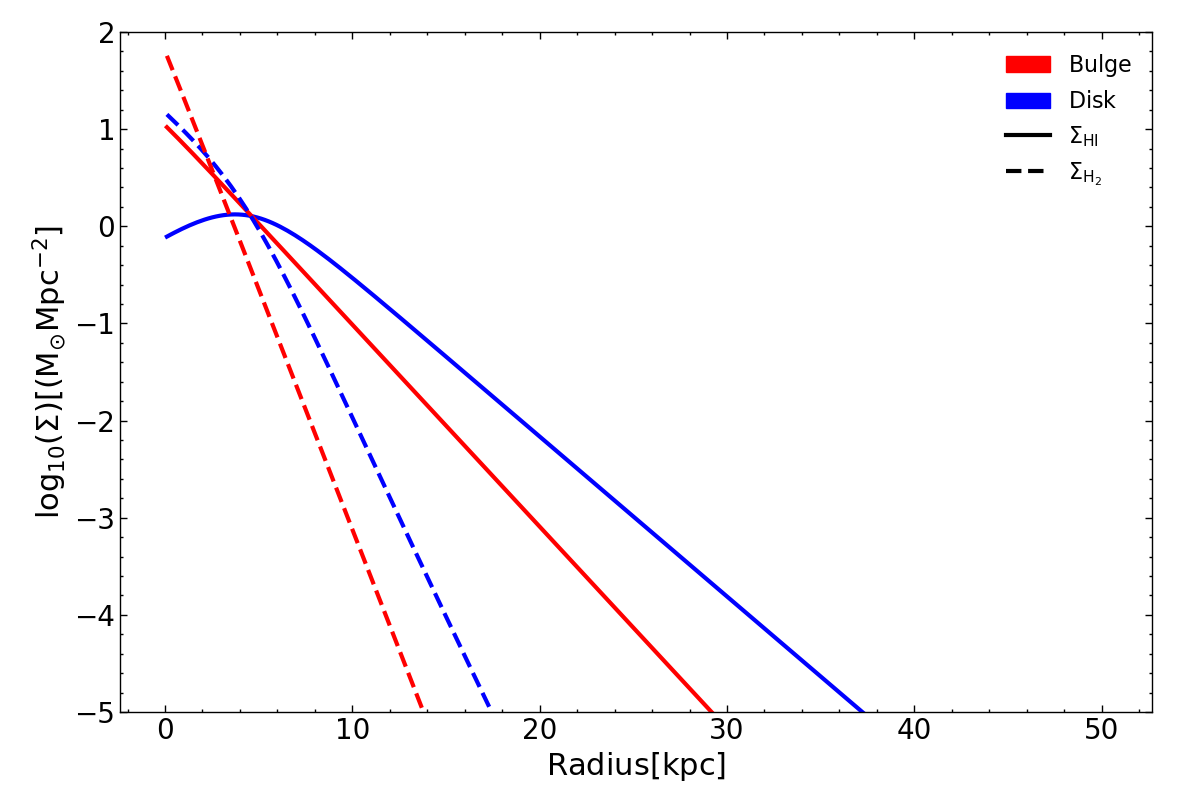}
	\caption{Surface density radial profiles of \hi\ in the disc and bulge, as labelled, for an example \shark\ galaxy, used to model the \hi\ emission lines. The solid and dashed lines represents the \hi\ and \hmol\ surface density of the galaxy, respectively. As it can be seen, there is a presence of \hi\ in the bulge of the galaxy, which drops down steeply in the beginning, but the \hi\ in the disc extends much further, and dominates beyond $\gtrsim 4$~kpc. There is a significant amount of \hmol\ present in the bulge, though it declines much more rapidly than the extended \hi\ disc.} 
	
	\label{fig:SAM-SurfaceProfiles}
\end{figure}

Figure~\ref{fig:SAM-SurfaceProfiles} shows the radial surface density profile for an example galaxy in \shark\, with a stellar and \hi\ mass of $10^9$ \M\ and $10^8$ \M, respectively. The inner radius is dominated by \hmol, with \hi\ forming a core there. The latter is due to the saturation of \hi\ at high column densities, above which the gas is converted into \hmol. The sum of both gas components is exponential, however, the individual ones can deviate from that assumption. \hi\ typically dominates at the outer radius. 

Previous work by \citet{Obreschkow2009} and \citet{Obreschkow2013} assumed the total gas disc to have an exponential profile with a scale length that was larger than the stellar one by a factor $>1$. They determined the \hi/\hmol\ ratio locally in post-processing using the \citet{Blitz2006} model, with updated empirical parameters obtained from THINGS \citep[The HI Nearby Galaxy Survey][]{Walter2008-THINGS}. Thus, our work improves on this by (i) allowing the \hi\ to have a more complex profile, such as the example of Figure~\ref{fig:SAM-SurfaceProfiles}, though still axisymmetric, and (ii) by calculating the multi-phase nature of galaxies self-consistently within the galaxy formation calculation. The latter directly impacts galaxy evolution as stars can only form from molecular hydrogen in \shark. In our model, \hi\ can also exist in the bulges of galaxies, which in general allows the models to reproduce the observed gas content of early-type galaxies \citep{Lagos2014,Serra-2010,Lagos2018Shark:Formation}.

\subsection{Circular velocity profile}
\label{sec:velocity_profile}

The circular velocity profiles are constructed following \citet{Obreschkow2009}, which we briefly describe in this section.  We assume 
a Navarro-Frenk-White (NFW; \citealp{Navarro1995-Xray}) halo radial profile, which describes the DM halo density profiles not as isothermal (i.e. $\rho \propto r^{-2}$) but with a radially varying logarithmic slope 

\begin{equation}
    \rho_{\rm halo}(r)\ = \ \rho_0\left[(r/r_{\rm s})(1 \ + \ r/r_s)^2\right]^{-1},
    \label{eq:NFW-halo}
\end{equation}

\noindent where $\rho_0$ is a normalization factor and $r_s$ is the characteristic scale radius of the halo (where the profile has a logarithmic slope of $-2$). The virial radius, $r_{\rm vir}$\, is calculated using the virial velocity of the haloes, $V_{\rm vir}$, following the relation,

\begin{equation}
    r_{\rm vir}\ =\ \frac{GM_{\rm vir}}{V_{\rm vir}^2},
    \label{eq:Virial-radius-calculation}
\end{equation}

\noindent where $M_{\rm vir}$ is the virial mass of the halo. Here, we define the virial mass as the mass enclosed within the halo when the overdensity is $200$ times that of critical density. The scale radius, $r_s$, is defined as $r_s\ = \ r_{\rm vir}/c_{\rm halo}$, where $c_{\rm halo}$ is the concentration parameter, which in \shark\ is estimated using the \citet{Duffy2008} relation. 

For a spherical halo, the circular velocity profile will be $V_{c}^{\rm halo^2} \ = \ \frac{GM_{\rm halo}(r)}{r}$, where $M_{\rm halo}(r)$ is the mass enclosed within the radius $r$. Therefore, the circular velocity profile of the halo is,

\begin{equation}
    V_{c}^{\rm halo^2}(x) = \left(\frac{GM_{\rm vir}}{r_{\rm vir}}\right) \ \times \ \frac{{\rm ln}(1+c_{\rm halo}x) - \frac{c_{\rm halo}x}{1 + c_{\rm halo}x}}{x \left[{\rm ln}(1+c_{\rm halo}) - \frac{c_{\rm halo}}{1 + c_{\rm halo}}\right]},
    \label{eq:velocity_halo}
\end{equation}

\noindent where $x\ \equiv \ r/r_{\rm vir}$. For larger radii, the circular halo velocity approaches the point mass velocity profile $V_{c}^{\rm halo^2} \ \approx \ GM_{\rm vir}/r$.


For the velocity profile of the disc, we use the stellar and gas surface densities calculated with \shark\. Stellar and gas surface density profiles are assumed to follow an exponential form with a distinct half mass radius for stellar and gas components. We calculate velocity profiles for stars and gas separately and then combine them to give $V_{\rm c}^{\rm disc}$. Following \citet{Obreschkow2009}, we define the circular velocity for the stellar disc, $V_{\rm c}^{\rm \star,disc}$, as 
\begin{equation}
\begin{multlined}
V_{\rm c}^{\rm \star,disc^2}(x) \approx  \frac{G\,M^{\rm \star,disc}}{r_{\rm vir}}\  \times \\ 
\frac{c_{\rm \star,disc} + 4.8\,c_{\rm \star,disc}\,{\rm exp}[-0.35\,c_{\rm \star,disc}\,x - 3.5/(c_{\rm \star,disc}\,x)]}{c_{\rm \star,disc}\,x + \left(c_{\rm \star,disc}\,x\right)^{-2} + 2\left(c_{\rm \star,disc}\,x\right)^{-1/2}},
\label{eq:velocity_disc}
\end{multlined}
\end{equation}

\noindent where $c_{\rm \star,disc} \equiv r_{\rm vir}/r_{\rm s,disc}$ is the stellar disc concentration parameter, where $r_{\rm s,disc} = r_{\rm \star,disc}/1.67$ is the scale radius of the stellar disc. $M^{\rm \star,disc}$ is the total mass of the stellar disc. We then calculate the contribution to the circular velocity from gas, $V_{\rm c}^{\rm gas}$, which we also describe as an exponential disc, and thus can be calculated as, 

\begin{equation}
\begin{multlined}
V_{\rm c}^{\rm gas^{2}}(x) \approx \frac{GM^{\rm gas}}{r_{\rm vir}}\ \times \\ \frac{c_{\rm gas} + 4.8c_{\rm gas}exp[-0.35c_{\rm gas}x - 3.5/(c_{\rm gas}x)]}{c_{\rm gas}x + (c_{\rm gas}x)^{-2} + 2(c_{\rm gas}x)^{-1/2}},
\label{eq:velocity_HI}
\end{multlined}
\end{equation} 

\noindent where $c_{\rm gas} \equiv r_{\rm vir}/r_{\rm s, gas}$ is the concentration parameter for the gas disc, where $r_{\rm s,gas} = r_{\rm gas}/1.67$. $M^{\rm gas}$ is the total cold gas mass (atomic plus molecular) of the galaxy. 

We note that Eqs.~\ref{eq:velocity_disc} and \ref{eq:velocity_HI} are an approximate solution for an exponential profile provided by \citet{Obreschkow2009a}. 

We describe bulges as spherical structures following a density profile according to the Plummer Model \citep{Plummer1911},

\begin{equation}
    \rho_{\rm bulge}(r) \approx \frac{3M^{\rm bulge}}{4 \pi r_{\rm Plummer}^3} \left[ 1 + \left( \frac{r}{r_{\rm Plummer}}\right)^2\right]^{-5/2},
    \label{eq:density_profile_bulge}
\end{equation}

\noindent with $r_{\rm Plummer} \ \approx \  1.7 r_{\rm bulge}$, and  $r_{\rm bulge}$ is the half-mass radius of the bulge. The contribution to the total circular velocity profile by the bulge is thus follows, 
\begin{equation}
V_{\rm c}^{\rm bulge^{2}}(x) = \frac{GM^{\rm bulge}}{r_{\rm vir}} \times \frac{(c_{\rm bulge}x)^{2} c_{\rm bulge} }{[ 1 + (c_{\rm bulge}x^{2})]^{3/2}}
\label{eq:velocity_bulge}
\end{equation}

\noindent where $c_{\rm bulge} \equiv r_{\rm vir}/r_{\rm s,bulge}$ is the bulge concentration parameter, where $r_{\rm s, bulge} = r_{\rm bulge}/1.67$. Unlike the $V_{\rm c}^{\rm disc}$ calculation, where we calculate gas and stellar terms separately, we assume gas and stars within the bulge to follow the same profile with the same scale radius when computing $V_{\rm c}^{\rm bulge}$; we combine their masses and calculate a single bulge contribution to the circular velocity profile. The latter was done as during the development of this model, we noted that the bulge gas and stellar radius were generally very similar and so we simply combined stellar and gas masses and used only the stellar bulge radius for our calculations.  

Now that we have all our components calculated, we can estimate the total circular velocity profile, $V_{\rm c}$ as,

\begin{equation}
    V_{\rm c}^2(x) \ = \ V_{\rm c}^{\rm halo^2}(x) + V_{\rm c}^{\rm \star,disc^{2}}(x) + V_{\rm c}^{\rm gas^{2}}(x) + V_{\rm c}^{\rm bulge^{2}}(x),
\label{eq:circular_velocity}
\end{equation}

\noindent which we use to construct the \hi\ emission line profiles.

\begin{figure}
	\includegraphics[trim=0mm 5mm 0mm 0mm,clip,width=\linewidth]{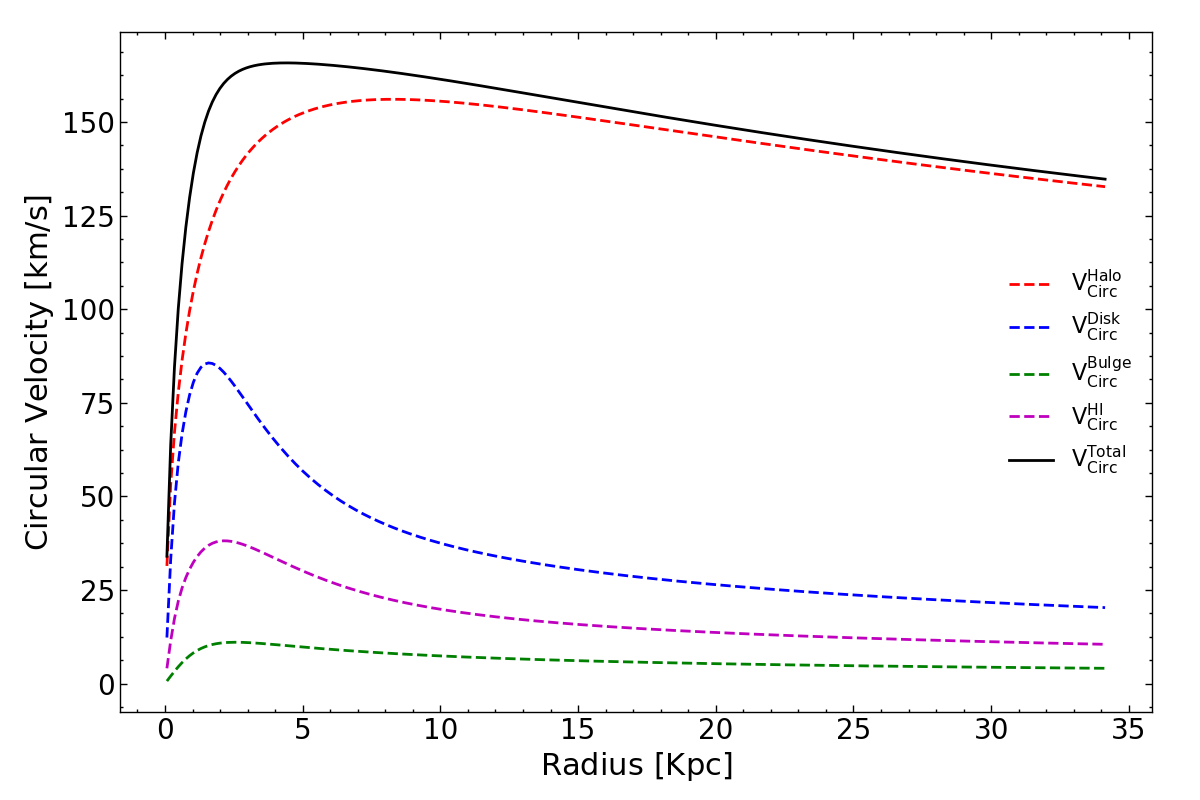}
	\caption{Radial circular velocity profile of the same galaxy showed in Figure~\ref{fig:SAM-SurfaceProfiles} (solid line), highlighting the contribution of all the components: stellar and gaseous disc, bulge and halo of the galaxy of, as labelled (see $\S$~\ref{sec:velocity_profile} for details). {The velocity profile of this galaxy is dominated by DM at all radii.}}
	\label{fig:SAM-VelocityProfile}
\end{figure}

\begin{figure}
	\includegraphics[trim=0mm 4mm 0mm 4mm,clip,width=\linewidth,height=7cm]{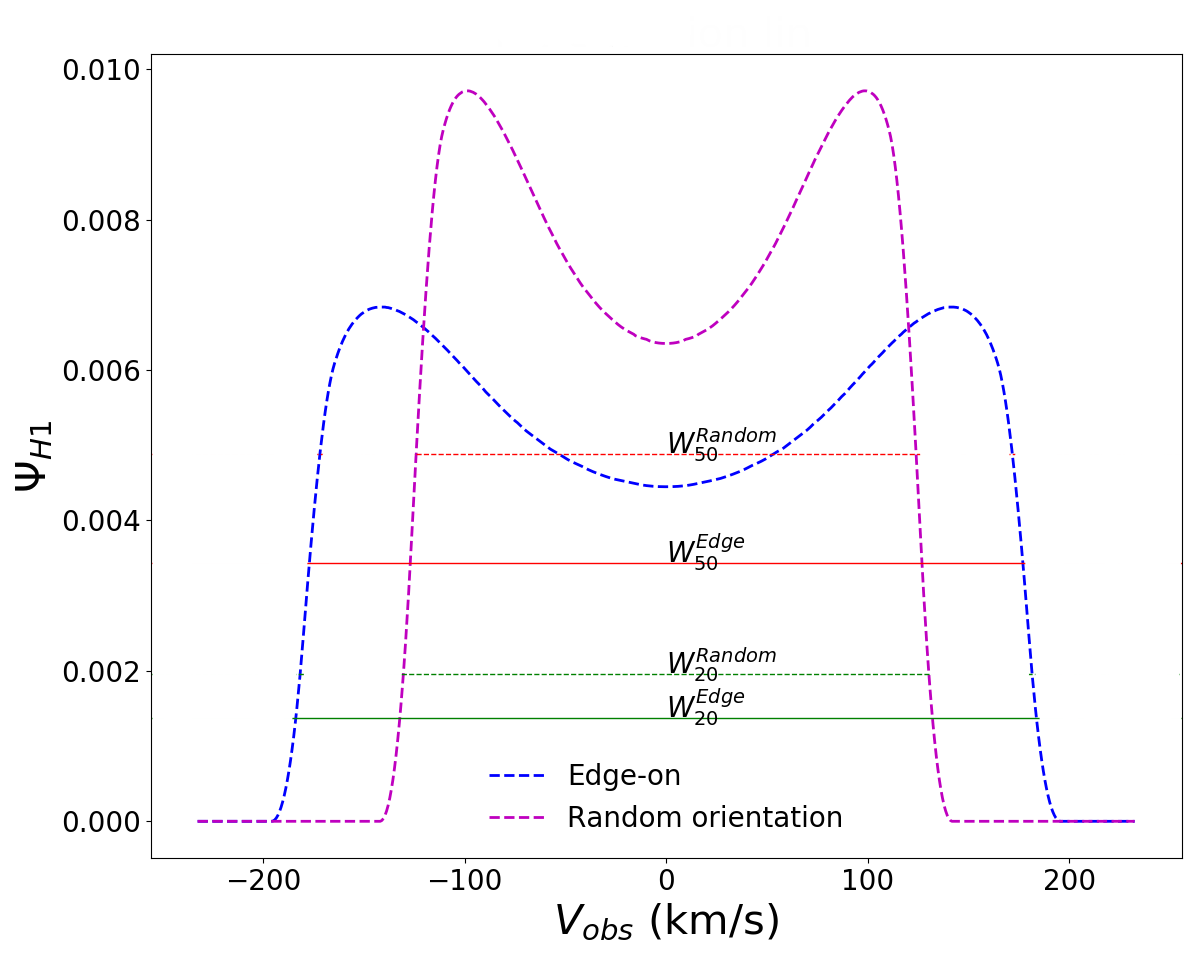}
	\caption{Normalised \hi\ emission line profile for the same example galaxy of Figs.~\ref{fig:SAM-SurfaceProfiles}~and~\ref{fig:SAM-VelocityProfile}, with edge-on and \textbf{intrinsic inclination of the randomly selected galaxy (in this case, $\rm cos\Theta \sim 60^\circ$), as labelled}. The two top and the two bottom horizontal lines mark the \wfifty\ and \wtwenty\ of the two orientations respectively. \wfifty\ and \wtwenty\ are maximal at edge-on orientations.}
	\label{fig:SAM-EmissionLines}
\end{figure}


\subsection{Emission Line Profile}
\label{sec:emissionline}

To construct the \hi\ emission line associated with any circular velocity profile, we consider the line profile of a flat ring with constant circular velocity $V_{\rm c}$ and a normalized flux. 

After imposing the normalization condition $\int dV_{\rm obs} \tilde{\psi}(V_{\rm obs}) \equiv 1$, the edge-on line profile of a ring is, 

\begin{equation}
\tilde{\psi}(V_{\rm obs}, V_{\rm c})  =
\begin{cases}
\frac{1}{\pi\sqrt{V_{\rm c}^{2}-V_{\rm obs}^{2}}} \hskip 3em \text{if} |V_{\rm obs}| < V_{\rm c} \\
0,\hskip 6.5em \text{otherwise.}
\end{cases}            
\end{equation}

\noindent This profile diverges as $|V_{\rm obs}| \rightarrow V_{\rm c}$, but the resulting singularity is smoothed by introducing a constant velocity dispersion for gas of $\sigma_{\rm gas} = 10{\rm km}\,{\rm s^{-1}}$ throughout the disc, which mimics the effect of random \hi\ motions. This assumption is supported by observations of the gas velocity dispersion seen in the nearby galaxies \citep{Leroy2008}. The smoothed normalized velocity profile is then given by

\begin{equation}
\psi(V_{\rm obs}, V_{\rm c}) = \frac{\sigma^{-1}}{\sqrt{2\pi}} \int dV\  \text{exp}\left[\frac{(V_{\rm obs} - V_{\rm c})^{2}}{-2\sigma^{2}}\right]\  \tilde{\psi}(V_{\rm obs}, V_{\rm c}). 
\end{equation}

\noindent From the edge-on line profile $\psi(V_{\rm obs}, V_{\rm c})$ of a single ring and the surface density of atomic hydrogen, $\Sigma_{\rm HI}$, which has been calculated as described in $\S$~\ref{sec:gasprofile}, we can construct the edge-on profile of the \hi\ emission line for the entire \hi\ disc, by using the following equation,

\begin{equation}
\Psi_{\rm HI}(V_{\rm obs}) = \frac{2\pi}{M_{\rm HI}} \int_{0}^{\infty} dr \  r \Sigma_{\rm HI}(r) \psi(V_{\rm obs}, V_{\rm c}(r)).
\end{equation}

\noindent An example of the resulting \hi\ emission lines is shown in Figure~\ref{fig:SAM-EmissionLines}, where we can see the signature double-horned profile. We include the effect of inclinations by using the inclination provided by \textsc{stingray} for every galaxy in the lightcone. 

To construct the \hi\ emission lines we assume a constant \hi\ velocity dispersion. Observations have found the latter to be remarkably constant, with values typically ranging from $8-12$\kms\ \citep{Leroy2008}, and approximately independent of galaxy properties. This has been suggested to be caused by thermal motions setting the \hi\ velocity dispersion, and the \hi\ abundance being largely dominated by the warm, neutral interstellar medium. Hence, we decide to keep this value constant, but note that increasing (decreasing) $\sigma_{\rm gas}$ has an effect of slightly increasing (decreasing) the number of low $W_{\rm 50}$ galaxies, $\lesssim 40$\kms\ in Figure~\ref{fig:VelocityFunction}.

\subsection{Flux calculation}
\label{sec:flux_calculation}

The lines described in $\S$~\ref{sec:emissionline} are normalized, and so need to multiply by the integrated flux of the \hi\ line to approximate an observed \hi\ emission line, which we do by using the relation of \citet{Catinella2010},

\begin{equation}
    \frac{M_{\rm HI}}{\rm M_{\odot}} = \frac{2.356 \times 10^{5}}{1 + z} \left[\frac{d_{\rm L}(z)}{\rm Mpc}\right]^2 \left(\frac{\int \rm S d\Omega}{\rm Jy \ \rm kms^{-1}}\right);
\end{equation}

\noindent here $M_{\rm HI}$ is the \hi\ mass, $d_L(z)$ is the luminosity distance of the galaxy at redshift $z$, and $\int \rm S d\Omega$ is the integrated flux. The luminosity distance and redshift information were obtained from the ALFALFA lightcone produced in the $\S$~\ref{sec:Lightcone} and the \hi\ mass is directly output by \shark. 


\subsection{How well does the \textmd{\hi} velocity width trace \texorpdfstring{$V_{\rm max}$}?}
\label{sec:vmax_vs_widths}

\begin{figure*}
\includegraphics[trim=3mm 2mm 3mm 2mm,clip,width=0.8\linewidth]{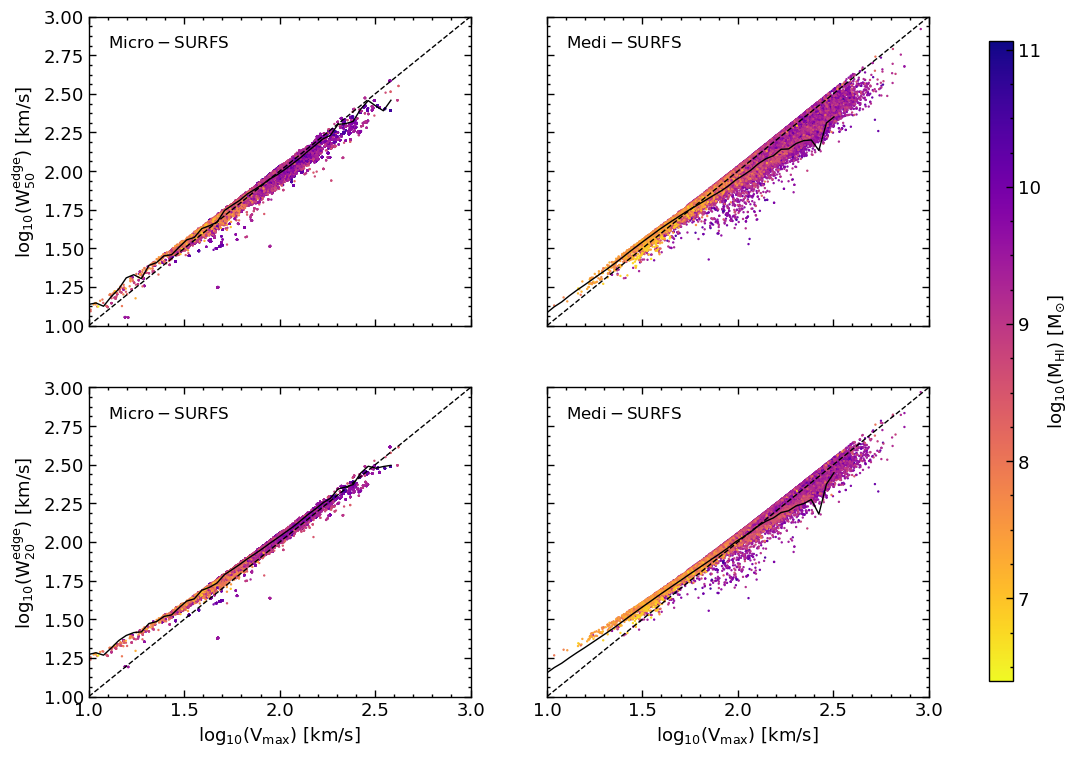}
\caption{Comparison of the intrinsic maximum circular velocities of \shark\ galaxies with that derived from our mock observations of the galaxies, using the width at $50$\% (top row) and $20$\% (bottom row) of the peak flux of the \hi\ emission lines of the simulated galaxies. The dashed and solid lines represent the 1:1 line and median of the values, respectively, with each scatter point being  an individual galaxy in the simulation, and coloured by their \hi\ mass, as shown in the colour bar at the right of the figure. A slight tendency to deviate up from the 1:1 relation is seen at $V_{\rm mas}\lesssim 30\,$\kms, which is caused by the fact that the \hi\ velocity dispersion and rotational velocity become comparable at such low velocities. As $W_{20}$ is measured at a lower level than $W_{50}$ it gets affected more by the dispersion than $W_{50}$.}

\label{fig:W20_vs_Vmax}
\end{figure*}


Figure~\ref{fig:W20_vs_Vmax} compares \vmax\ and the $50^{\rm th}$ percentile, \wfifty, and $20^{\rm th}$ percentile, \wtwenty, widths of the \hi\ emission lines in the case of edge-on orientations, for all galaxies in the ALFALFA lightcone (see $\S$\ref{sec:velocity_profile} for a description of \vmax); \wfifty\ and \wtwenty\ are widely used in the observations to estimate rotational velocities of galaxies.

Figure~\ref{fig:W20_vs_Vmax} shows that there is good agreement between the true maximum circular velocities and the simulated \hi\ \wfifty\ and \wtwenty\ at the higher velocity regime, $V_{\rm max}\gtrsim 100\,$\kms, but there are systematic deviations at lower velocities, $V_{\rm max}\lesssim 35\,$\kms. These deviations can be understood as the effect of non-circular motions modelled via the inclusion of the random \hi\ velocity component to the \hi\ emission lines. As stated in $\S$~\ref{sec:emissionline}, we incorporate a velocity dispersion of $10$ \kms\ throughout the \hi\ disc. When we reach the low velocity range ($\lesssim 35$ \kms), this velocity dispersion is comparable to these circular velocity of the disc and skews the \hi\ linewidths. We should also note that the direction of this skewness is the opposite to what \citet{Brooks2017} found in their cosmological hydrodynamical zoom simulations of dwarf to MW galaxies. In spite of this effect, however, we can recover the observed \hi\ velocity and mass distributions $\S$~\ref{sec:Reproducing ALFALFA}.


\subsection{\textmd{\hi} line profiles: Idealised models vs. Hydrodynamical simulations}
\label{sec:appendix_introduction}

As discussed in $\S$~\ref{sec:Modelling}, we assume profiles for our dark matter, gas and stellar components when modelling the \hi\ emission lines of all \shark\ galaxies. In addition, we also assume axis-symmetry that leads to perfect double-horned \hi\ emission line profiles for our \shark\ galaxies. Observations show that asymmetries in the \hi\  emission line profiles are common \citep{Catinella2009} and hence we would like to test how much our assumptions affect our ability to predict a distribution of \wfifty\ and \wtwenty. 

With this aim, we use a suite of 13 dwarf and 2 Milky-Way sized galaxies from the APOSTLE cosmological hydrodynamical simulations suite \citep{Sawala2016ThePuzzles} as a test-bed, and use the \textsc{MARTINI} \citep{Oman2019} software to produce \hi\  emission lines for all these galaxies (see Appendix~\ref{sec:appendix} for details). We find that our idealised model reproduces very well the \wtwenty\  measurements of the APOSTLE simulations. However, the \wfifty\ measurements show more discrepancies driven by the asymmetry of the \hi\ emission lines in the APOSTLE simulations. These deviations are typically within $\approx 25$\% in the case of dwarf galaxies $V_{\rm max}\lesssim 100$~\kms, while being larger for the $2$ Milky-Way galaxies. Because we are interested primarily in the dwarf regime, we conclude that our idealised HI emission line model produces a good enough representation of dwarf galaxies even in hydrodynamical simulations.

\section{Reproducing the \textmd{\hi} masses and velocities of observed galaxies in a \texorpdfstring{\lcdm}\ \  framework}
\label{sec:Comparisons}

We compare \shark\ predictions with \hi\ observations to highlight the conclusions one could draw in such case. We then go onto comparing our simulated ALFALFA survey with the real one and discuss our findings.


\subsection{A raw comparison between \textmd{\shark} and the observed \textmd{\hi}\ masses and velocities of galaxies}
\label{sec:Raw comparison}
\begin{figure*}
  \includegraphics[trim=5mm 5mm 0mm 3mm, clip,width=0.45\linewidth,height=8cm]{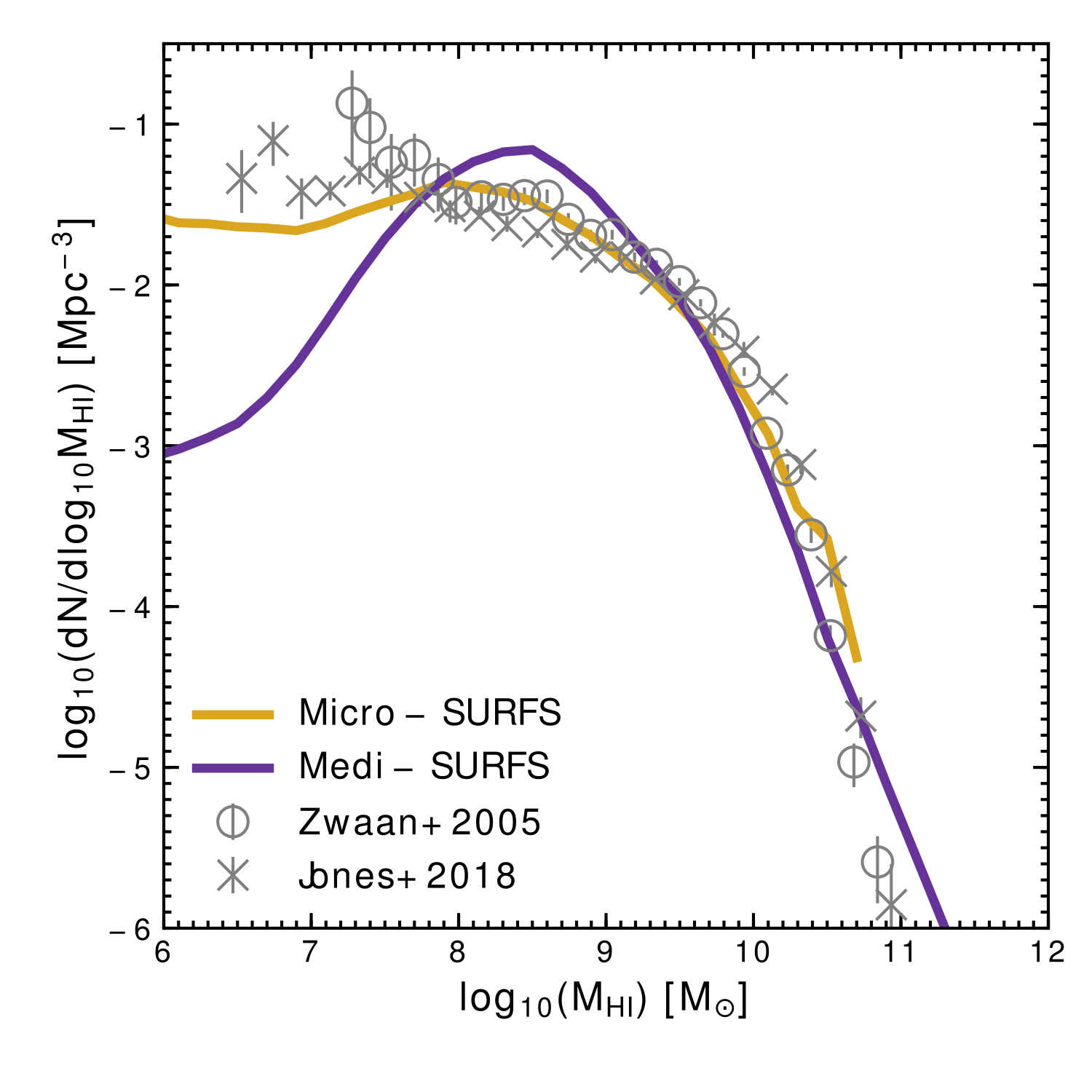}
  \includegraphics[width=0.45\linewidth,height=8cm]{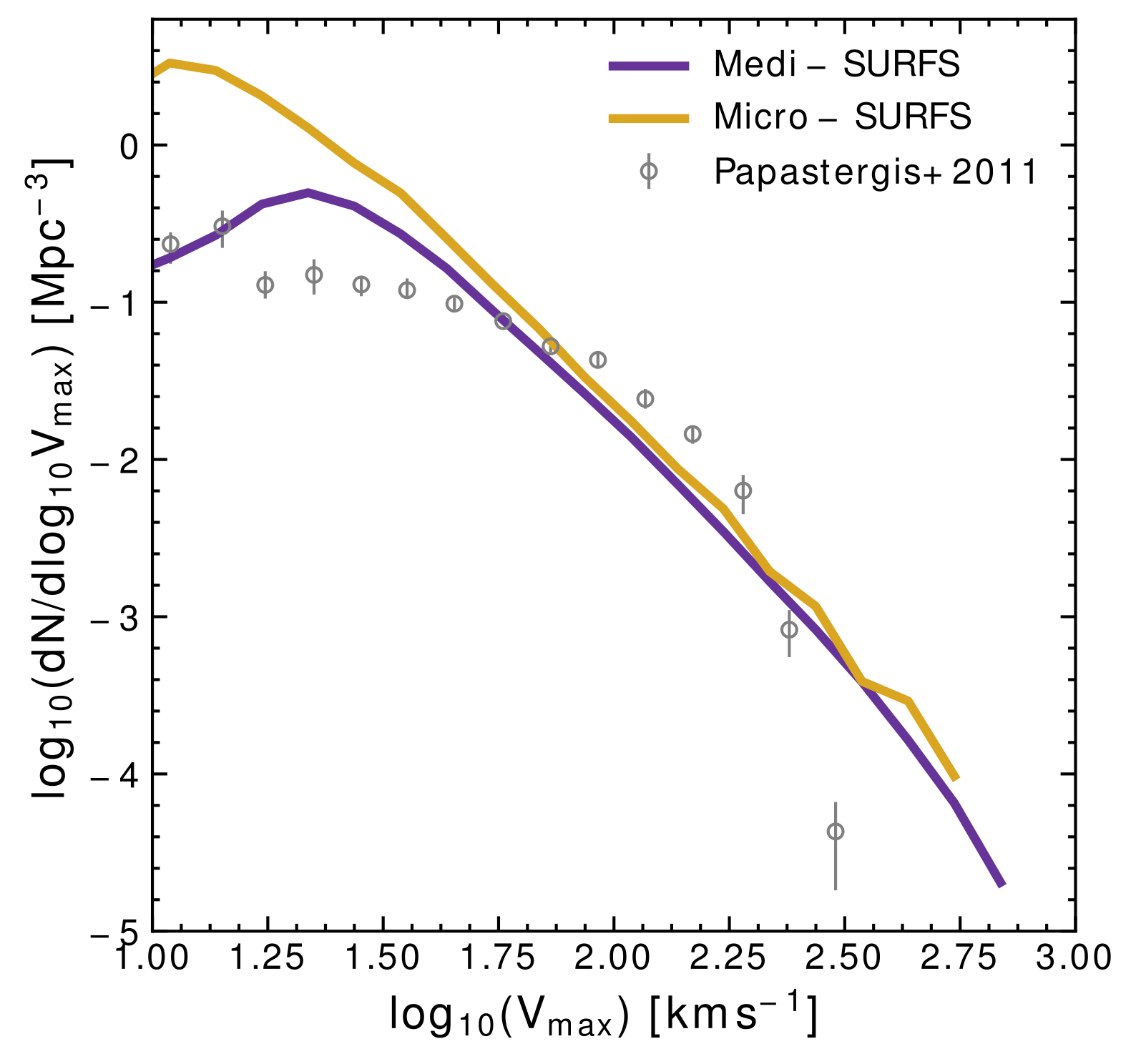}
\caption{The \hi\ mass function (left panel) and \hi\ velocity function (right panel) of all the \shark\ galaxies at $z = 0$, produced using the medi-\surfs\ and micro-\surfs, as labelled in each panel. We also show as symbols the observational estimates from \citet{Zwaan2005,Jones2018} in the case of the \hi\ mass function, and from \citet{Papastergis2011} for the \hi\ velocity function. There is good agreement between the \shark\ and the observations of the \hi\ mass function, while there is a clear tension with the observations of the \hi\ velocity function at $V_{\rm mas}\lesssim 100\,$\kms}.

\label{fig:Raw Comaprison}
 \end{figure*}
 
 
\begin{figure}
    \centering
    \includegraphics[trim=1mm 3mm 3mm 4mm, clip,width=\linewidth,height=8cm]{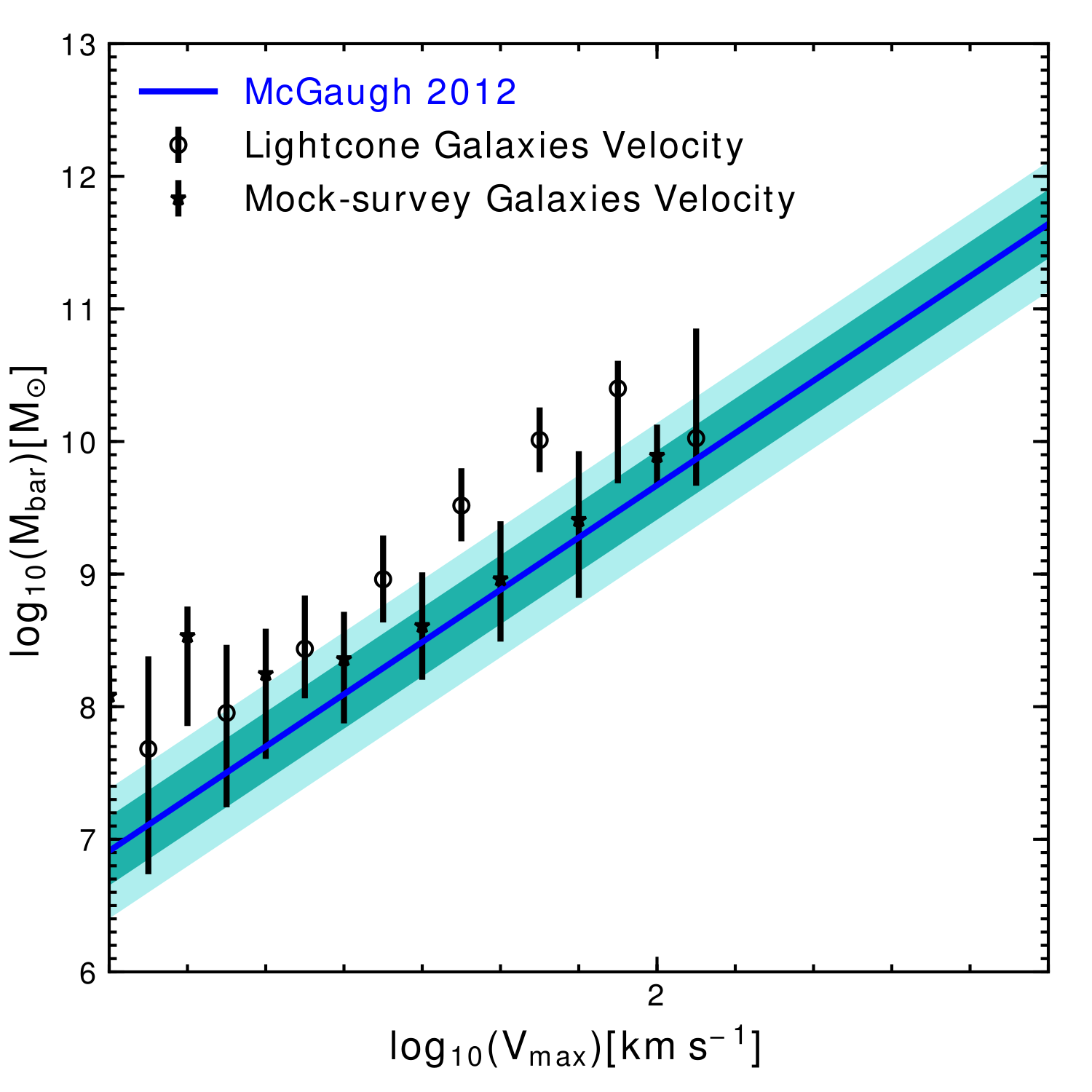}
    \caption{The baryonic Tully-Fisher relation of all the galaxies in the lightcone compared to those that we flag as ``ALFALFA-selected'' in the lightcone. We also show the best fit to the observed relation from \citet{McGaugh2011}. We show the results from the micro-\surfs\ box only as there was little difference in the values from medi-\surfs. The figure shows that the entire galaxy population follows a Tully-Fisher relation in tension with the observations, while the more fair comparison with the ``ALFALFA-selected'' simulated galaxies shows much better agreement, showing that \shark\ galaxies reproduce the Tully-Fisher relation very well.
    }
    \label{fig:Tully Fisher}
\end{figure}

The traditional way in which simulations are compared to observations is by taking the predicted galaxy population in the simulated box and comparing directly with derived properties of galaxies in observational surveys. The drawback of such an approach is that there may be important selection biases that are not taken into consideration. This could lead us to conclude that the simulation fails to reproduce an observable when in fact it reflects a mismatch in the different selections and biases that are present in simulation and observational data. This hampers interpretation of the shortcomings of simulations and our understanding of galaxy formation. 

In this context, we examine the raw \shark\ predictions with the derived ALFALFA \hi\ mass and velocity functions, which should illustrate the importance of accounting for selection effects. We do the comparison using both the micro-\surfs\ and medi-\surfs\ (see $\S$~\ref{sec:Galaxy Catalogue} for details) simulations, and perform a raw comparison with ALFALFA. This assumes that observations are able to sample an unbiased portion of the galaxy population across the probed dynamic range and hence, a reliable volume correction can be applied to take the observed distributions to convert them into functions.

In the left panel of Figure~\ref{fig:Raw Comaprison}, we compare the \hi\ mass function at $z=0$ that we derive from \shark, running over the two simulation boxes described in $\S$~\ref{tab:sims}, with the observed \hi\  mass function at $z=0$ from \citet{Jones2018} and \citet{Zwaan2005}, and find overall agreement between the predictions and observations. Micro-\surfs\ agrees better with the observations across the whole dynamic range of masses observed, while medi-\surfs\ agrees well with the observations at $M_{\rm HI}\gtrsim 10^9\,\rm M_{\odot}$, while deviating at lower \hi\ masses. This difference is simply a resolution effect, in which the haloes that host central galaxies with $M_{\rm HI} \lesssim 10^9\,\rm M_{\odot}$ are not well resolved in medi-\surfs, but they are in micro-\surfs. The median halo mass for central galaxies of $M_{\rm HI}\lesssim 10^9\,\rm M_{\odot}$ is $M_{\rm Halo} \lesssim 10^{11.4}\,\rm M_{\odot}$ in the medi-\surfs, which would comprise of $\sim 1100$ particles in them. On the other hand, micro-\surfs\ has a similar median halo mass for central galaxies below $M_{\rm \hi}\lesssim 10^9\,\rm M_{\odot}$, but because of better mass resolution such halo masses ae made of $\approx 6000$ particles, and so is able to better resolve the haloes over the dwarf galaxy mass range. The agreement between \shark\ and observations is not surprising because \citet{Lagos2018Shark:Formation} used the \hi\ mass function as a guide to find a suitable set of values for the free parameters in \shark.  

In the right panel of Figure~\ref{fig:Raw Comaprison}, we show the comparison between the $40$\% ALFALFA data release global \hi\ velocity function at $z=0$ as calculated by \citet{Papastergis2011} and the ``raw'' \hi\  velocity functions of the circular velocities of the galaxies at $z=0$ in \shark, again for our two simulations, medi-\surfs\ and micro-\surfs. This allows us to determine whether or not \shark\ over-predicts the number of low dynamical mass systems as reported in \citet{Zavala2009}, \citet{Schneider2016}, \citet{Papastergis2011} and \citet{Obreschkow2013}. We find that more galaxies are predicted than are observed by more than an order of magnitude at circular velocities $<100\,\rm km\,s^{-1}$. The peak of the velocity function for micro-\surfs\ is shifted towards a lower velocity ($\sim 20$\kms)  due to its higher mass resolution, which enables us to better sample the low dynamical mass galaxies at the cost of producing a smaller number of massive galaxies. The latter is due to the smaller volume. {This problem is remedied by including medi-\surfs, which allows us to access much larger cosmological volumes and hence higher dynamical masses. The downside is that its resolution is coarser and hence does not go down to the low halo masses that we have access to with micro-\surfs. The two simulations in combination allow us to fully sample the velocity and \hi\ mass range of interest, $\approx 20$~\kms\ to $800$~\kms. We confirm previous results that have reported an over-abundance of low-dynamical mass galaxies in \lcdm\ compared to observations, even after accounting for the complexity of how galaxies populate haloes through the modelling of \shark.}

Because we are investigating the masses and velocities of galaxies, it is natural to extend the comparison to the Tully-Fisher relation \citep{Tully-Fisher}, which is an empirical relation between the optical luminosity and the \wfifty\ of \hi\ emission lines. The Tully-Fisher relation has been used to place tight constraints on galaxy formation models and is used as a test for the robustness of those models (e.g. \citealp{Fontanot-2017-TFR-example}). \citet{McGaugh2011} extended the classic Tully-Fisher relation to the baryonic Tully-Fisher relation (BTFR), which relates the total baryonic mass of galaxies (gas plus stars) with the observed rotational velocities. In Figure~\ref{fig:Tully Fisher}, we compare the predicted BTFR of all disc-dominated (bulge-to-total ratio $< 0.5$) \shark\ galaxies (open symbols) with the observed BTFR of \citep{McGaugh2011}. Here, we only show the micro-\surfs~because the medi-\surfs~results are similar, albeit lacking the lowest $V_{\rm circ}$ galaxies. We find that the simulated galaxies tend to be $\approx 0.2-0.3$~dex more \hi\ massive at fixed circular velocity compared to observations. If instead we use the edge-on \hi\ \wfifty\ of galaxies that are present in our mock survey, we find that they follow the BTFR more closely. This result further strengthens our confidence in that the \hi\ \wfifty\ measurements done in this study are a closer representation of the observed \hi\ \wfifty\ than raw circular velocity.

\begin{figure*}
\begin{center}
	\includegraphics[trim=0mm 3mm 2mm 0mm, clip,width=0.8\linewidth]{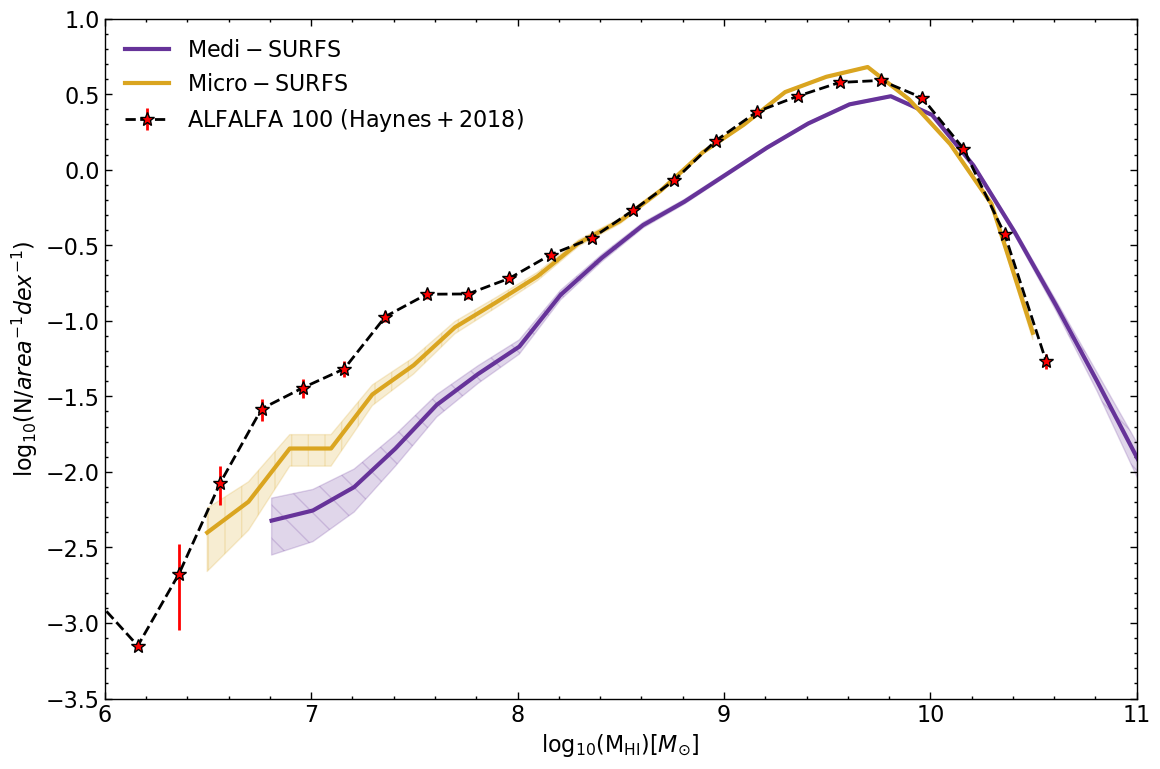}
	\caption{Comparison of the \hi\ mass distribution as obtained from our mock ALFALFA survey with the observations of \citet{Haynes2018TheCatalog}. The purple and yellow solid lines represent the results of the lightcones constructed with \shark, using the medi-\surfs\ and micro-\surfs\ $N$-body simulations, respectively. The shaded region is representative of the poisson noise in the data. Our mock survey's \hi\ mass distribution, in both resolution boxes, is in reasonable agreement with the observations.}
	\label{fig:MassFunction}
\end{center}
\end{figure*}

\begin{figure*}
\begin{center}
	\includegraphics[trim=0mm 3mm 2mm 0mm, clip,width=0.8\linewidth]{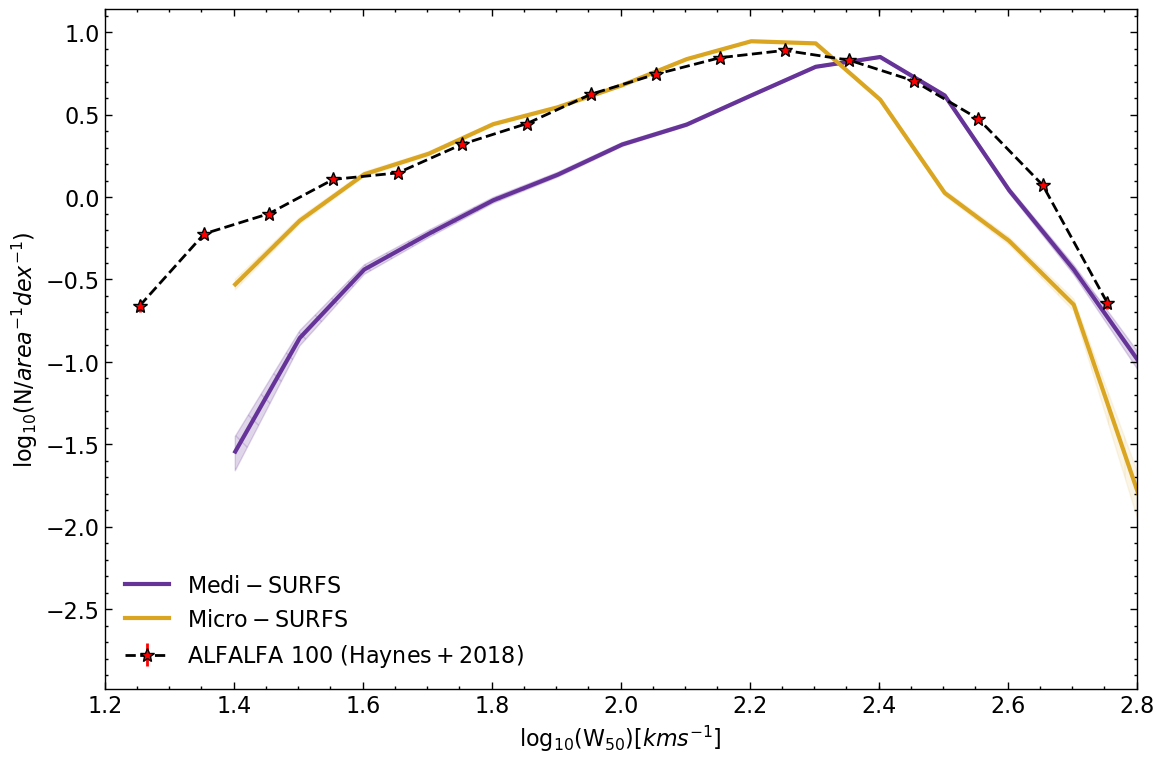}
	\caption{The \hi\ velocity distribution obtained by our mock ALFALFA survey, with the purple and yellow solid lines representing the \shark\ model run over the medi- and micro-\surfs\ simulations, respectively, with the shaded regions representing the poisson noise. Because micro-\surfs\ has a higher resolution than medi-\surfs, it traces the lower velocity end better, while the medi-\surfs\ is able to track down the galaxies at higher velocity end. By combining the results from these two boxes and applying the selection function of ALFALFA, we are able to obtain a velocity function that is in agreement with the observations.}
	\label{fig:VelocityFunction}
\end{center}
\end{figure*}



\begin{figure*}

\includegraphics[trim=0mm 3mm 2mm 1mm,clip,width=\linewidth]{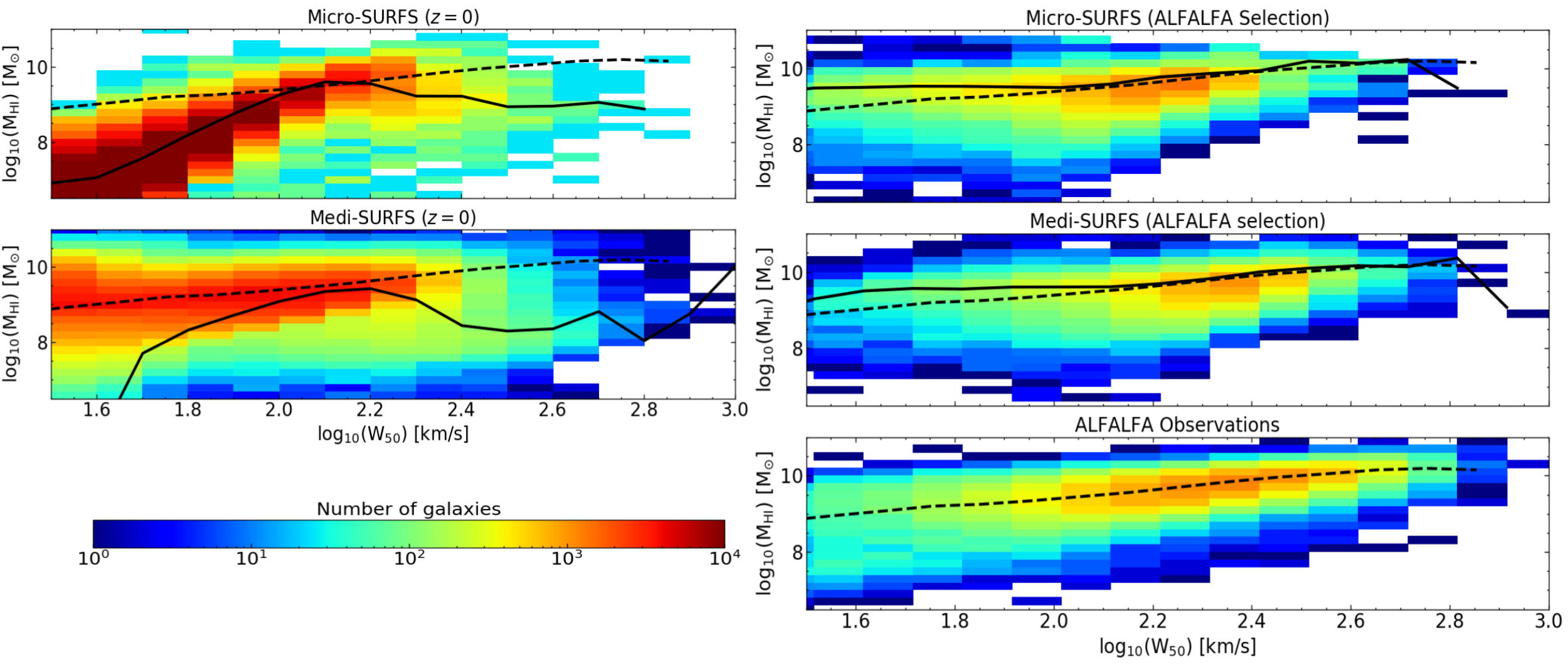}

\caption{2D histograms showing the number of galaxies in the plane of \hi\ mass and \wfifty\ for the \shark\ galaxies obtained by running the model in the medi-\surfs\ and micro-\surfs, as labelled. The left-hand panels show {\it all} the galaxies in the simulation at $z=0$, which we scale accordingly to match the volume of ALFALFA, whereas the right-hand panels show only the galaxies that are comply with the ALFALFA selection in our mock survey. The bottom, right-hand panel shows the actual observed \hi\ mass-\wfifty\ relation of the ALFALFA survey as released in \citet{Haynes2018TheCatalog}. The colour bar indicates the number of galaxies present in each bin. Solid lines show the running median for that respective panel whereas the dashed line is the running median for the ALFALFA observations. Most galaxies in the model are below the ALFALFA selection criterion which is why the relations look so different between the left and right panels. Anyhow, the similarity to the actual observations gives us the confidence that we are detecting similar galaxies in our mock survey.} 

\label{fig:2DHist}
\end{figure*}



\begin{figure*}
\minipage{0.5\linewidth}
  \includegraphics[width=\linewidth]{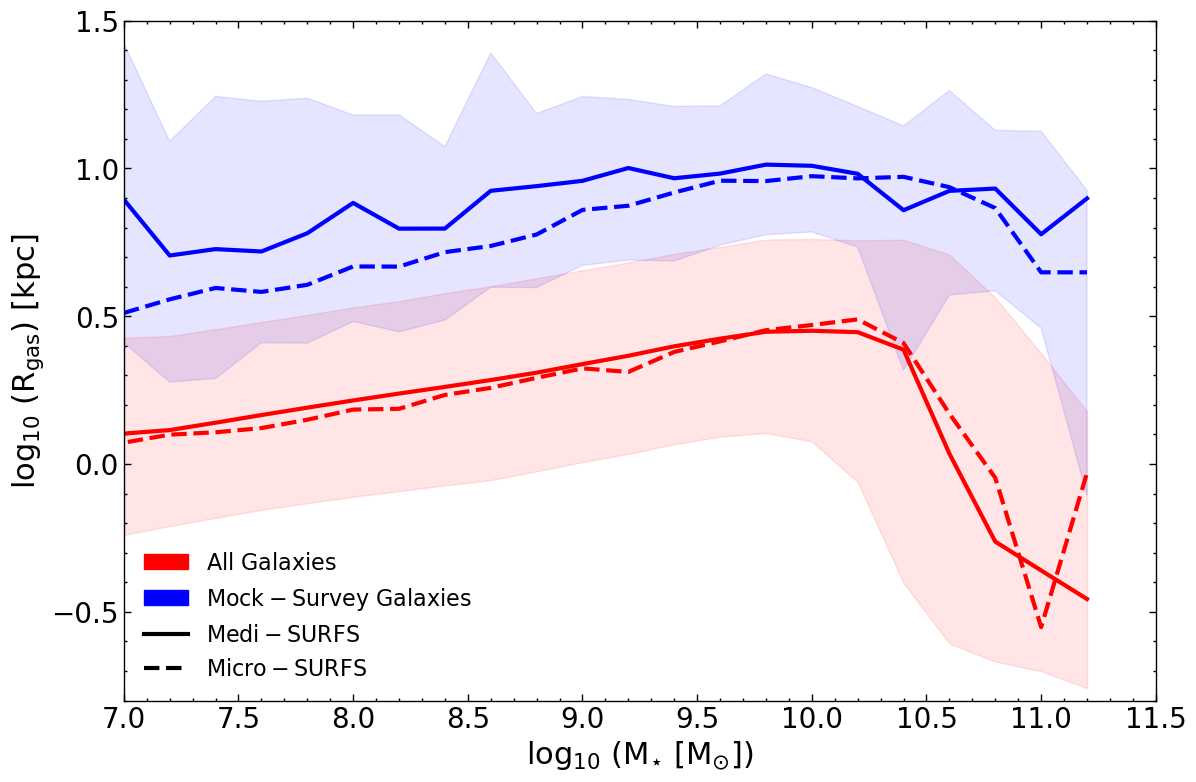}
  
\endminipage\hfill
\minipage{0.5\linewidth}
  \includegraphics[width=\linewidth]{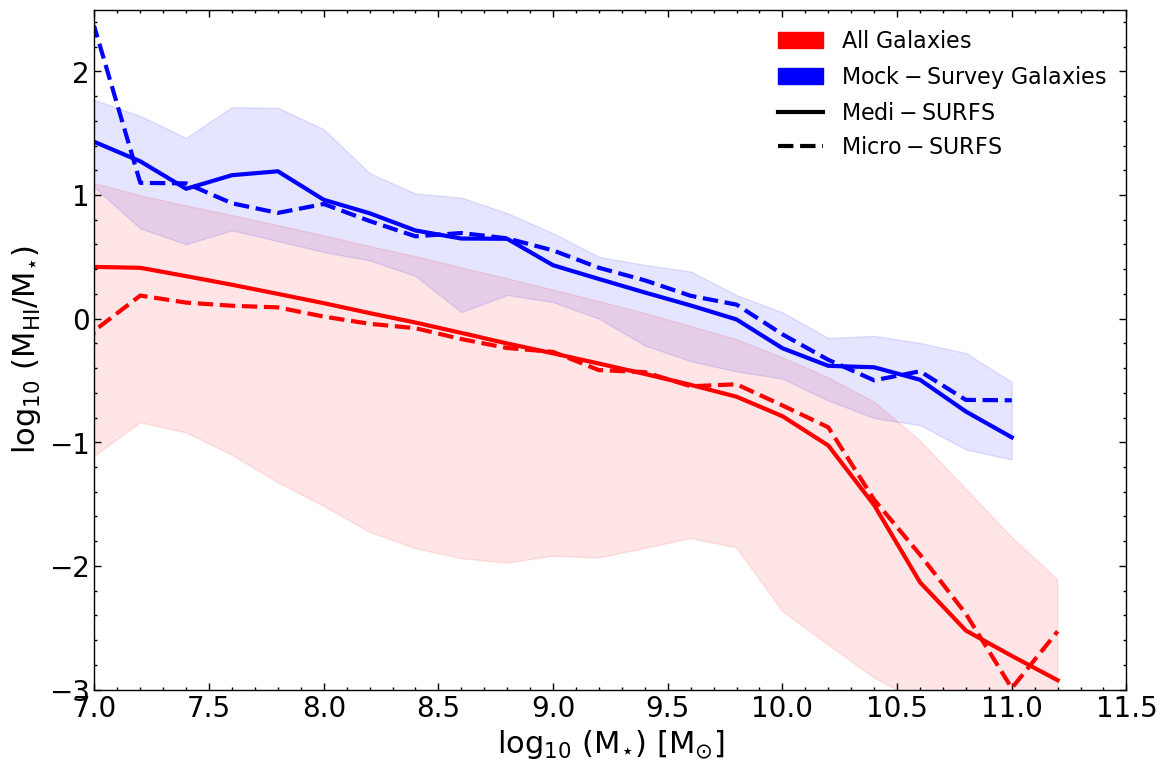}
  
\endminipage
\caption{Half-gas mass disc radius (left panel) and \hi-to-stellar mass ratio (right) as a function of stellar mass of the galaxies at $z=0$ in \shark. The lines and colours represent our two simulations medi- and micro-\surfs, as labelled. Shaded regions show the $16^{\rm th}-84^{\rm th}$ percentiles. For clarity, the latter are shown only for the medi-\surfs. A clear selection effect is seen as galaxies with larger gas discs and higher gas-to-star ratio are preferentially selected by ALFALFA.}
\label{fig:gas_disc_mhi}
\end{figure*}



\begin{figure*}
\minipage{0.5\linewidth}
  \includegraphics[width=\linewidth]{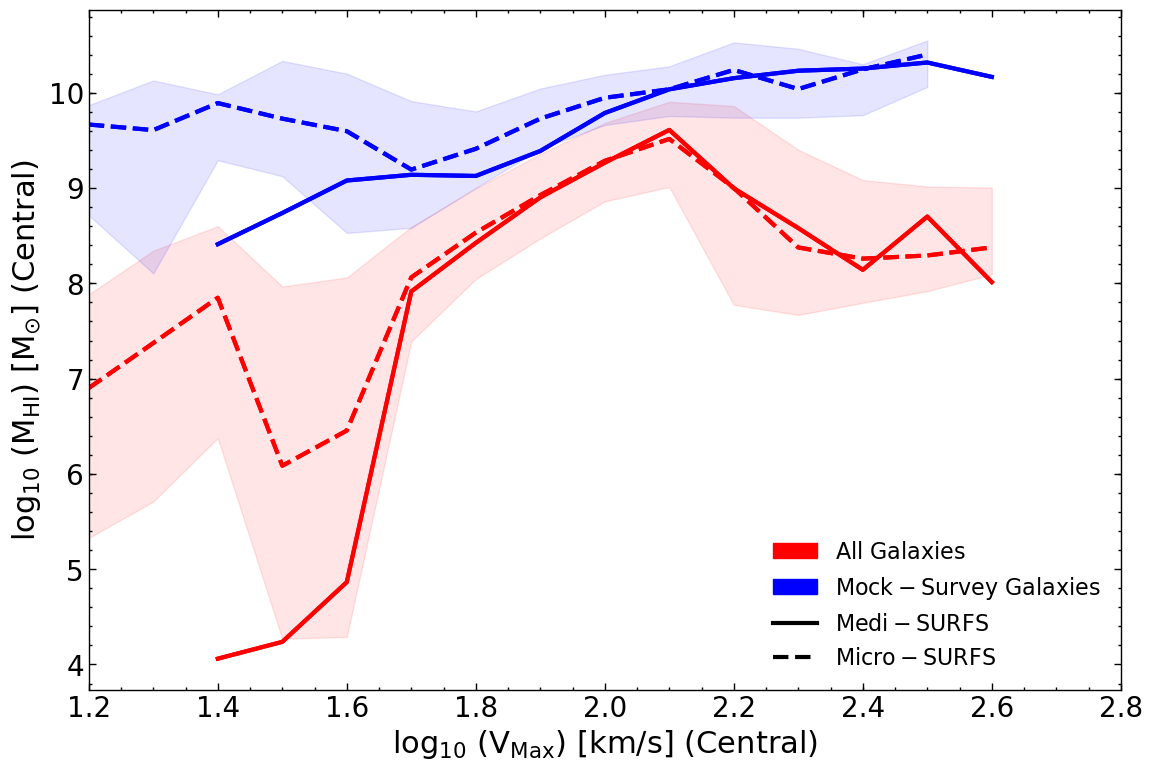}
  
\endminipage\hfill
\minipage{0.5\linewidth}
  \includegraphics[width=\linewidth]{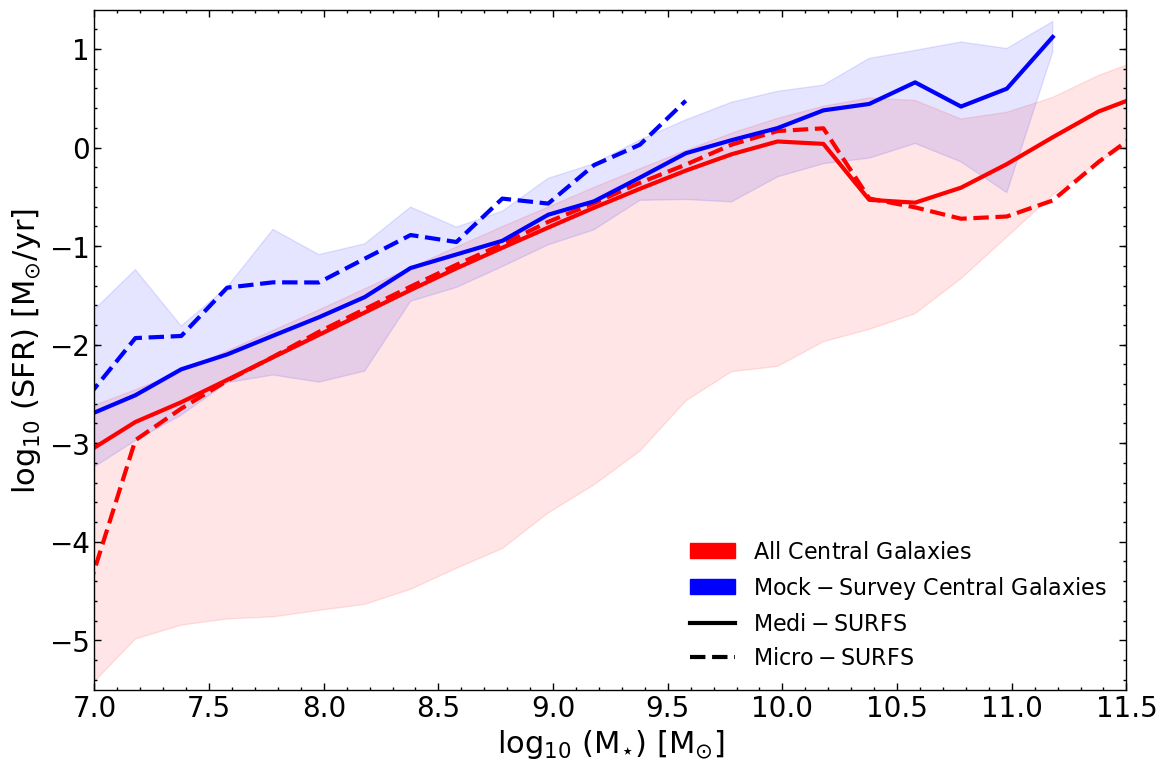}
  
\endminipage
\caption{\textit{Left}: \hi\ content of galaxies as a function of the maximum circular velocity of the galaxy (which is used as a proxy for dynamical mass). Due to the limited resolution of medi-\surfs, we only shown the latter down to ${\rm log}_{10}(V_{\rm max}/{\rm km\,s^{-1}})=1.4$. Resolution is the likely driver of the difference seen between medi- and micro-\surfs\ below  ${\rm log}_{10}(V_{\rm max}/{\rm km\,s^{-1}})\approx 1.7$. 
Here, we show the $16^{\rm th}-84^{\rm th}$ percentiles for micro-\surfs\ as it goes down to lower circular velocities.
\textit{Right}: As Figure~\ref{fig:gas_disc_mhi} but for the star-formation rate (SFR) as a function of the stellar mass. In both panels a clear bias is seen as the ALFALFA mock-survey is preferentially selecting galaxies with higher \hi\ content, albeit a smaller bias is seen for the star formation rate.}
\label{fig:vdm_sfr_mhi}
\end{figure*}



\subsection{A mock-to-real comparison between \textmd{\shark} and  ALFALFA}
\label{sec:Reproducing ALFALFA}

The Arecibo Legacy Fast ALFA (ALFALFA) survey is a `blind' \hi\ survey that has mapped nearly $7000$ deg\textsuperscript{2} area in the velocity range $-2000 < cz < 18,000$ \kms, where $c$ is the speed of light and $z$ is the redshift. The survey has identified $\sim 31,500$ extragalactic \hi\ line sources \citep{Haynes2018TheCatalog}. The detection limit of the survey as described by \citet{Papastergis2011} is a function of the integrated \hi\ line flux,$S_{\rm int,lim}$,  and velocity width  $S_{\rm int,lim}/{\rm Jy}~{\rm kms}^{-1} = 0.06~(W_{50}^{0.51}/{\rm km s}^{-1})$.

For our analysis, we apply the same selection of \citet{Papastergis2011} to our lightcones (see $\S$~\ref{sec:Lightcone} for details) to  select ALFALFA-like galaxies; this results in our mock ``ALFALFA'' survey. We remind the reader that our lightcone has the same survey area and redshift coverage as ALFALFA. We also apply beam confusion to the lightcone prior to applying the selection criterion above. 

We construct the \hi\ mass distribution from the released catalogue of \citet{Haynes2018TheCatalog}, and present this as number per unit deg$^2$. The resulting observed distribution is shown in Figure~\ref{fig:MassFunction} as symbols. We perform the same measurement in our mock ALFALFA survey (one for each \surfs\ simulation being used here), which we also show in Figure~\ref{fig:MassFunction}. We find that there is very good agreement between the simulated and observed \hi\ mass distributions, which is particularly striking for the lightcone based on micro-\surfs. This is not surprising, because Figure~\ref{fig:MassFunction} shows that the predicted \hi\ mass function agrees well with the measurements of \citet{Jones2018}. There is a slight tension between \hi\ masses of $10^7$\M\ and $10^8$\M, where \shark\ predicts a slightly lower number of galaxies. \citet{Lagos2018Shark:Formation} showed that the abundance of galaxies below the break of the \hi\ mass function was very sensitive to the adopted parameters in the photo-ionisation model. Lower velocity thresholds, below which haloes are not allowed to cool gas to mimic the impact of a UV background, has the effect of producing a higher abundance of low \hi\ mass galaxies (see their Appendix~A). 

In this work we do not attempt to calibrate \shark\ to reproduce the low-mass end of the \hi\ mass function but simply to show how our default model performs compared to \hi\ observations, and to put constraints on the magnitude of the discrepancy (if any) between the predictions and the observations of \hi\ masses and velocity widths.\\

We now turn our attention to the \hi\ \wfifty\ distribution. We take the \hi\ \wfifty\ measurements from \citet{Haynes2018TheCatalog} (which are as observed, and hence there is no attempt to correct by inclination effects), and construct the \hi\ \wfifty\ distribution (shown as symbols in Figure~\ref{fig:VelocityFunction}). We also take our modelled \hi\ \wfifty\ (assuming the {\sc stingray} inclinations for our simulated galaxies) and construct the \hi\ \wfifty\ distribution for those that pass the ALFALFA selection criterion for our two lightcones created running \shark\ on the medi- and micro-\surfs\ (lines in Figure~\ref{fig:VelocityFunction}, as labelled). We find that the model and the observations agree remarkably well. We remind the reader that the observationally derived \hi\ velocity function and the $V_{\rm max}$ function of \shark\ displayed differences of factor $\gtrsim 20$ at velocities $\lesssim 30$ \kms\ (see Figure~\ref{fig:Raw Comaprison}), while in Figure~\ref{fig:VelocityFunction}, differences are $\lesssim 50$\%. In other words, the ``missing dwarf galaxy problem'' is not evident. Using the medi- and micro-\surfs\ allow us to probe the entire range of the observations with the micro-\surfs\ simulation probing the lower velocity end $\lesssim 30$\kms, while the medi-\surfs\ allows us to improve significantly the statistics at the high  \hi\ \wfifty\ end $\gtrsim 100$\kms. With \shark\ applied to these two simulations, we are able to reproduce the observed \hi\ \wfifty\ distribution. The large differences seen between Figure~\ref{fig:Raw Comaprison} and Figure~\ref{fig:VelocityFunction} suggests that there are important selection biases which cannot be easily corrected in the process of taking the observed \hi\ \wfifty\ distribution and inferring from there an \hi\ \wfifty\ function, which prevent us from making a one-to-one comparison between the predicted $V_{\rm max}$ function from DMO simulations and observations. This highlights the fact that building lightcones to reproduce observational surveys is essential to tackle this problem, and, in their absence, erroneous conclusions could be drawn. 

We have so far shown that \shark\ produces galaxies with the correct \hi\ mass and \wfifty\ distributions, but that does not necessarily mean that galaxies of a given \hi\ mass have the right \hi\ \wfifty. To test this, Figure~\ref{fig:2DHist} shows 2D histograms of galaxies in the \hi\ mass-\wfifty\ plane. The left panel shows all the galaxies in the simulation at $z=0$, whose numbers are scaled accordingly to match the ALFALFA volume, whereas the right panel shows the galaxies which pass the ALFALFA selection criterion applied to our lightcones. We also show the same 2D histograms of galaxies for the real ALFALFA survey in the bottom, right panel of Figure~\ref{fig:2DHist}. Going from left to right panels of Figure~\ref{fig:2DHist} show that the majority of galaxies that were originally present in simulation box do not satisfy the ALFALFA selection. 
Large differences are seen between the 2D distributions of the galaxies in the $z=0$ simulated boxes and the mock ALFALFA lightcones.
Most of the galaxies in both micro- and medi-\surfs\ with masses $M_{\rm HI}\lesssim 10^9\,\rm M_{\odot}$ are selected out, producing a narrower relation between \hi\ mass and \wfifty\ than the one followed by the underlying population of simulated galaxies. Our simulated ALFALFA lightcone reproduces well the observed \hi\ mass and \wfifty\ relation of ALFALFA. However, there is some tension in the medians as \shark\ tends to produce $0.1-0.4$~dex too much HI mass at ${\rm log}_{10}(W_{\rm 50}/{\rm km \,s^{-1}})\lesssim 2.1$. This difference is also seen in  Figure~\ref{fig:MassFunction}, as the number of galaxies in the simulations is less than the observed one in the regime of ${\rm M_{\rm HI}} \lesssim 10^{8} {\rm M_{\odot}}$.

In Figures~\ref{fig:gas_disc_mhi} and \ref{fig:vdm_sfr_mhi}, we show the biases the selection criterion of ALFALFA introduces in the galaxy population; in other words, how do ALFALFA-like galaxy properties compare to the underlying galaxy population? In both figures, the red and the blue colours represent all galaxies in the lightcone (prior to any selection) and the ALFALFA mock-survey galaxies (after applying the ALFALFA selection), respectively.

Figure~\ref{fig:gas_disc_mhi} and the right panel of Figure~\ref{fig:vdm_sfr_mhi} show the half-gas mass disc radius, \hi-to-stellar mass ratio and star-formation rate (SFR) as a function of the galaxy stellar mass, for all galaxies in \shark\ and selected by the ALFALFA criteria (i.e. those that make up the distributions of Figures~\ref{fig:MassFunction}~and~\ref{fig:VelocityFunction}). The left panel of Figure~\ref{fig:vdm_sfr_mhi} compares the \hi\ content of the galaxies with its dark matter halo circular velocity, for the sub-sample of central galaxies in both \shark\ and in those selected as ALFALFA-like. When comparing the gas radii (see \textit{left} panel in Figure~\ref{fig:gas_disc_mhi}), we see that the median of the ALFALFA mock survey galaxies is always higher than the overall median of galaxies in \shark\ (i.e. the underlying galaxy population), with our simulated ALFALFA galaxies having a half-gas mass radius of the disc $\approx 0.5-0.7$~dex larger than \shark\ galaxies of the same stellar mass at $M_{\star}\lesssim 10^{10.3}$\M. A drop in the half-gas mass radii of galaxies at stellar masses higher than $10^{10.3}$ \M is seen for the overall median of the \shark\ galaxies (red). The latter is due to  this mass range being dominated by passive elliptical galaxies which tend to be gas poor. This drop is not seen in the median of the ALFALFA mock survey galaxies (blue), thus showing that ALFALFA preferentially picks out gas-rich galaxies, avoiding early-type galaxies that are affected by AGN feedback. This preference is clear when we compare the $\frac{M_{\rm HI}}{M_{\star}}$ ratio for both observed and all galaxies in the \shark\ (see right panel in Figure~\ref{fig:gas_disc_mhi}), with the mock ALFALFA survey galaxies, which continue to be systematically gas richer than the overall median, even at the dwarf galaxy regime.  

We also see a strong preference for gas-rich galaxies when we compare the maximum circular velocity of central galaxies with their \hi\ content (see left panel in  Figure~\ref{fig:vdm_sfr_mhi}), with the mock observed galaxies median (blue) staying in the range of $10^8 \rm M_{\odot} \lesssim M_{\rm HI} \lesssim 10^{10} M_{\odot}$, even when the overall median (red) is orders of magnitude below ($M_{\rm HI}\sim 10^6-10^8\,\rm M_{\odot}$). Even though both ALFALFA and our mock ALFALFA survey detect galaxies with \hi\ content as low as $10^6$\M, the number of those detections are fairly low ($\sim 20-30$ galaxies), making the higher \hi\ mass galaxies more dominant and skewing the median towards those values even at the low circular velocity end. 

When analysing the overall central galaxy population, there is a clear peak in the $M_{\rm HI}-V_{\rm max}$ relation, which is related to the peak of the baryon collapse efficiency in galaxies (e.g. \citealt{Eckert2017}). \citet{Baugh2018-PMillennium} using the GALFORM  semi-analytic model of galaxy formation \citep{Cole_2000,Lacey2016,Lagos2014} also found a sharp break in the \hi\ mass-halo mass relation at $\gtrsim 10^{11.5}\,\rm M_{\odot}$. 
This is the approximate halo mass scale at which AGN feedback starts to suppress gas cooling in both models, leading to the decline in \hi\ mass. The width and prominence of the peak is therefore expected to be very sensitive to the AGN feedback model and hence a useful relation to constrain from observations.

When comparing the star formation rate (SFR) with the stellar mass (see right panel in  Figure~\ref{fig:vdm_sfr_mhi}), we see only a small tendency for the ALFALFA mock survey galaxies to have slightly higher SFRs than the underlying galaxy population, again across the whole stellar mass range studied here. The most probable reason for this effect is that in \shark\ the SFR is calculated from the \hmol\ content of the galaxies, which in turn depends on the total gas mass and radius. Because gas \textbf{masses} are larger in the ALFALFA mock survey galaxies compared to the underlying population, that tends to drive a smaller H$_2$/HI ratio, which is why the SFRs in Figure~\ref{fig:vdm_sfr_mhi} are close to the median of \shark\ despite the higher \hi\ abundance in Figure~\ref{fig:gas_disc_mhi}. The main sequence of star formation of the entire sample of lightcone galaxies shows a clear break at $\sim 10^{10}$\M, driven by the mass above which AGN feedback starts to be important (typically overcoming the gas cooling luminosity). This break is not seen in the ALFALFA mock survey galaxies, showing the strong bias against gas poor, low star-forming galaxies. 

These biases are to be expected because ALFALFA is a blind survey and is limited by the integrated \hi\ flux and velocity width, which in turn depends on the \hi\ mass content of galaxies. What is unexpected is that these biases are important even at the dwarf galaxy regime, where most galaxies are star-forming and gas-rich; our ALFALFA mock survey galaxies are more gas-rich and more star-forming. This also raises concerns regarding how  best to correct for the galaxies that are not detected by ALFALFA, and how to account for the fact that the observed population is not representative even at the dwarf galaxy regime. Thus, we can see that selection bias plays a very important role in our understanding of the intrinsic galaxy properties and are crucial even at dwarf galaxy scales.  


\subsection{Implications for \texorpdfstring{\lcdm}~~and comparisons with previous studies}
\label{sec:Discussion}

\citet{Brooks2017} used a suite of $33$ cosmological zoom hydrodynamical simulations, covering a wide dynamic range from dwarfs to MW-like galaxies, and suggested that the dearth of observed galaxies with low circular velocities was caused by the \hi\ line-width (used as the dynamical mass tracer) not tracing the full potential well in dwarf galaxies. The reason for this was because in their simulated dwarf galaxies, the bulk of HI is in the rising part of the rotation curve, which means that the integrated HI line width does not reflect the maximum circular velocity of the galaxy. This results in a relation between the effective circular velocity of \hi\ ($V_{\rm HI}=W_{\rm 50}/2$ for a galaxy observed edge-on) and the maximum circular velocity which significantly deviates from the 1:1 relation at the dwarf galaxy regime, in a way that in the latter $V_{\rm HI}$ is much smaller than $V_{\rm max}$. By applying the relation $V_{\rm HI}-V_{\rm max}$ obtained from their zoom simulations to the dark matter halos of a large cosmological volume, DM-only simulation, they were able to reproduce the observed galaxy velocity function. This therefore offers an attractive solution to the tension seen in Figure~\ref{fig:Raw Comaprison}, which is also supported by the fact that there have been reports from observations in some nearby dwarfs that the bulk of HI is indeed in the rising part of the rotation curve e.g. \citet{Catinella2005,Swaters2009,Oman2019}.

\citet{Maccio2016} arrived at a similar conclusion, but using mock-observed galaxies from the NIHAO simulations suite\citep[a suite of 100 cosmological hydrodynamical simulations zooms, again covering a wide dynamic range from dwarfs to MW-like galaxies][]{NIHAO2015}. They obtained similar deviations of the $V_{\rm HI}-V_{\rm max}$ relation from the 1:1 line at the dwarf galaxy regime as Brooks et al. Two reasons were given by \citet{Maccio2016} to explain this, one was again the fact that \hi\ is not extended enough to reach the flat part of the rotation curve, and the second was that the non-circular motions of the gas seem to become significant at the dwarf galaxy regime (also seen in other cosmological zoom simulations; e.g. \citealt{Oman2019}). Despite this impressive progress, an important limitation remains. Both studies, \citet{Maccio2016} and \citet{Brooks2017}, assume their suite of simulated galaxies to be representative of all the galaxies of the same $V_{\rm max}$. The main question is then whether $33$ or $100$ galaxies is sufficient to make a statement about the main drivers of the tension seen in Figure~\ref{fig:Raw Comaprison}.

To address this question we turn to our ALFALFA lightcones and quantify the fraction of galaxies at two maximum circular velocities, $V_{\rm max}=100$~\kms\ and $V_{\rm max}=30$~\kms\ that would be selected by ALFALFA (given their selection criteria) in a fixed cosmological volume. These $V_{\rm max}$ values are chosen because the deviations of the $V_{\rm HI}-V_{\rm max}$ relation from the 1:1 line in \citet{Maccio2016,Brooks2017} appear at $V_{\rm max}\lesssim 100$~\kms. In \shark, we find that $\approx 22$\% of the galaxies with $V_{\rm max}=100$~\kms\ would be detectable by ALFALFA, while that number reduces to $\approx 1.4$\% for galaxies with $V_{\rm max}=30$~\kms. In the context of the simulated samples of \citet{Maccio2016,Brooks2017}, a few galaxies with $V_{\rm max}=100$\kms and $<1$ (or $\sim 0.462$) galaxy with $V_{\rm max}=30$~\kms\ would be detectable by ALFALFA. In addition, the small fraction of dwarf galaxies that would be detectable by ALFALFA is far from representative of the galaxies that have on average the same stellar or halo mass. This strongly argues for the need of large statistics to assess the tension between \lcdm\ and the observed galaxy velocity function of Figure~\ref{fig:Raw Comaprison}.

Our work therefore differs from previous ones in two fundamental ways. The first is that we use a statistically significant population of galaxies; with each simulated box having $\sim 1.3$ million galaxies, each of which have their own star formation, gas accretion and assembly histories, and so we are capable of simulating the entire ALFALFA survey volume. The second is that is that 
we obtain a $V_{\rm HI}-V_{\rm max}$ relation that is very close to the 1:1 line even at the dwarf galaxy regime. Hence, we are able to reproduce the observed \hi\ \wfifty\ distribution {\it without} the need to invoke significant deviations in the $V_{\rm HI}-V_{\rm max}$ relation. That is not to say these deviations do not exist but simply that {\it observations can be reproduced without them}. 
The fact that our model does not obtain the deviations discussed above is likely due to the simplistic physics that is inherent to semi-analytic models, which are much better captured with hydrodynamical simulations, and therefore likely reflects a limitation of our model. In the Appendix, we applied our idealized model to galaxies in the APOSTLE hydrodynamical simulation suite, and found that in dwarf galaxies our method overestimates \wfifty\ by $\approx 20-30$\%. If we were to correct out \wfifty\ distribution of Figure~\ref{fig:VelocityFunction} by these differences, our predicted number of dwarf galaxies would slightly decrease, making the number of dwarfs {\it smaller} than the observed one - indicating that the observed abundance of low \wfifty\ galaxies is very sensitive to baryon physics.


Our work suggests that the main effect in the apparent discrepancies between the predicted $V_{\rm max}$ function from DMO simulations and the recovered one from observations are selection effects, which are complex because of how non-linearly galaxy properties correlate with their halo properties. Hence, the \hi\ velocity distribution is not a cosmological test, but more appropriately a baryon physics test. This also strongly suggests that for a complete and unbiased understanding of \hi\ galaxy surveys, it is necessary to mock-observe our simulated galaxy population and compare with observations in a like-to-like fashion. 

\section{Conclusions}
\label{sec:conclusions}

The abundance of galaxies of different maximum circular velocities (the velocity function) is a fundamental prediction of our concurrent cosmological paradigm and hence, of uttermost important to test against observations. 
In this work, we have used the \shark\ semi-analytic galaxy formation model to simulate the ALFALFA \hi\ survey, the largest blind \hi\ survey to date, to investigate the well-known discrepancy between the observed and predicted galaxy \hi\ velocity function. Our goal was to determine whether this tension is a true failure of \lcdm, or simply a reflection of the complexity of baryon physics. 

We have presented how we model \hi\ emission lines in \shark\, taking into account halo, gas and stellar radial profiles of galaxies, and tested our idealised approach against more complex models derived from the cosmological hydrodynamical APOSTLE simulations by comparing our derived \hi\ line widths with theirs and find good agreement.
We used this new modelling to build a mock ALFALFA survey, and in the process, we combined simulation boxes spanning a range of mass resolutions and cosmological volumes, to ensure a good coverage over the full dynamical range probed by the observations. By applying the ALFALFA selection function to our simulated galaxies, we were able to recover the observed \hi\ velocity and mass distributions to within $30$\%, which shows that {\it a physically motivated model of galaxy formation in the \lcdm\ paradigm is able to reproduce the observed \hi\ velocity width distribution of galaxies}. We highlight that these are true predictions of our \shark\ model, as gas properties are a natural outcome of the model and were not included in fine tuning of the free parameters of the model.

Our key results can be summarised as follows - 

\begin{itemize}
    \item Survey selection plays a major role in explaining the discrepancy between  predictions and observations of the HI velocity function. We see an over-prediction of galaxies in the \hi\ velocity function of more than an order-of-magnitude at the low velocity end only when we make an ``out-of-the-box'' comparison of the predicted and observed galaxy populations, while a careful comparison accounting for the survey selection criteria reveals discrepancies of less than $50$\%. On applying the ALFALFA selection criteria, we get the desired \hi\ \wfifty\ distribution even at low circular velocities, alleviating the missing dwarf galaxy problem.
    
    \item Our predicted galaxy population agrees well with the observed \hi\ mass function. We compare the \hi-\wfifty\ 2D distribution obtained from the $100$\% data release of ALFALFA with our mock survey, and find agreement at an acceptable level. This strengthens our belief that the discrepancy between the predicted \hi\ velocity distribution with the observed one is due to the selection biases inherent in the survey.
    
    \item Previous simulations found that the effective HI velocity ($V_{\rm HI}=$\wfifty$/2$ for an edge-on galaxy) significantly underestimates $V_{\rm max}$, which has been invoked as a plausible explanation for the discrepancies described above in the velocity function. We find that our HI emission line modelling produces a $V_{\rm HI}-V_{\rm max}$ relation that is very close to the 1:1 line even at the dwarf galaxy regime. Despite this, we are able to reproduce the \hi\ \wfifty\ distributions; these deviations may still happen, but we argue that they are not necessary to reproduce the observed \hi\ \wfifty\ distribution. 
    
    \item A clear selection bias is seen when the mock is compared with the total galaxies that are presented in \shark, shown in Figures \ref{fig:gas_disc_mhi} and \ref{fig:vdm_sfr_mhi}. The mock ALFALFA survey is biased towards galaxies with a higher \hi\ gas content, larger \hi\ sizes and slightly higher SFRs. We find that at fixed $V_{\rm max}$ the mock ALFALFA galaxies are very strongly biased towards high \hi\ masses, with a difference in the typical \hi\ mass of up to two orders of magnitude at $V_{\rm max}\approx 30-50$\kms. This selection bias, in turn, affects our understanding of the distribution of galaxies in our local universe. Thus in order to fully understand galaxy evolution, a clear understanding of these biases is required.
    
    \item By comparing our simple model of HI emission lines with the more complex HI lines obtained in the cosmological hydrodynamical simulation APOSTLE, we find that \wtwenty\ is less affected by the asymmetry that is seen in the \hi\ emission lines than \wfifty, the more commonly used velocity estimator. Thus, robust observational measurements of \wtwenty\ would be extremely useful to constrain the simulations and uncover any tension with the simulations.
      
\end{itemize}

Our study suggests that the primary reason for the discrepancy between the \hi\ velocity function in observations and \lcdm\ simulations are selection effects in HI surveys, which are highly non-trivial to correct for. The latter is due to the fact that the typical galaxy with low circular velocity detected in ALFALFA is far from representative of galaxies of the same stellar or halo mass, particularly at $V_{\rm max}\lesssim 100$~\kms, according to our predictions. The observed \hi\ velocity distribution is therefore an excellent test for the baryon physics included in our cosmological galaxy formation models and simulations rather than a cosmological one.

A new generation of HI surveys is underway in telescopes such as The Australian Square Kilometer Array Pathfinder (ASKAP; \citealt{Johnston-2008-ASKAP}). Examples of those are the Widefield ASKAP L-band Legacy All-sky Blind surveY (WALLABY: \citealp{WALLABY_paper}) and the Deep Investigation of Neutral Gas Origins (DINGO: \citealp{Meyer-2009-DINGO}). The depth of these surveys will certainly lead to improvements over previous HI surveys; however, a careful consideration of systematic effects such as those described here will be necessary to make measurements that can be robustly compared with simulation predictions. Similarly, the exercise of simulating the selection effects of surveys to the detail presented here, will be equally important to identify the areas in which our understanding of galaxy formation and perhaps cosmology need improvement.


\section*{Acknowledgements}

We thank Tom Quinn, Kristine Spekkens, Robin Cook, Barbara Catinella, Aaron Ludlow, Bi-Qing For, Jesus Zavala, Joop Schaye and Marijn Franx  for constructive comments and useful discussions.
GC is funded by the MERAC Foundation, through the Postdoctoral Research Award of CL, and the University of Western Australia.
We also thank Aaron Robotham and Rodrigo Tobar for their contribution towards \surfs\ and \shark, and Mark Boulton for his IT help.
Parts of this research were carried out by the ARC Centre of Excellence for All Sky Astrophysics in 3 Dimensions (ASTRO 3D), through project number CE170100013. CL and PE are funded by ASTRO 3D. KO received support from VICI grant 016.130.338 of the Netherlands Foundation for Scientific Research (NWO). This work was supported by resources provided by the Pawsey Supercomputing Centre with funding from the Australian Government and the Government of 
Western Australia.




\bibliographystyle{mnras}
\bibliography{mnras_template}



\appendix
\section{Assessment of our HI emission line model against the APOSTLE cosmological hydrodynamical simulations}
\label{sec:appendix}
\begin{figure*}

    \begin{minipage}[10cm]{\textwidth}
        \includegraphics[scale = 0.3]{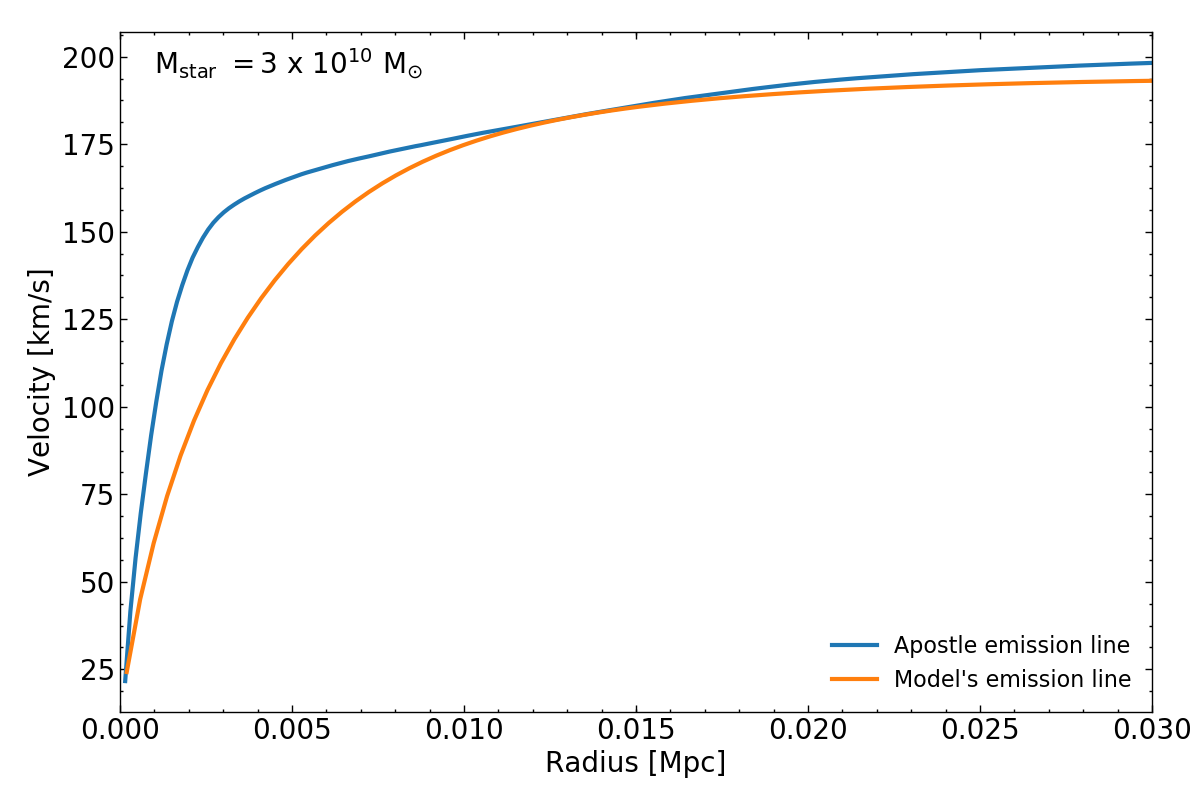}
        \hfill
        \includegraphics[scale = 0.3]{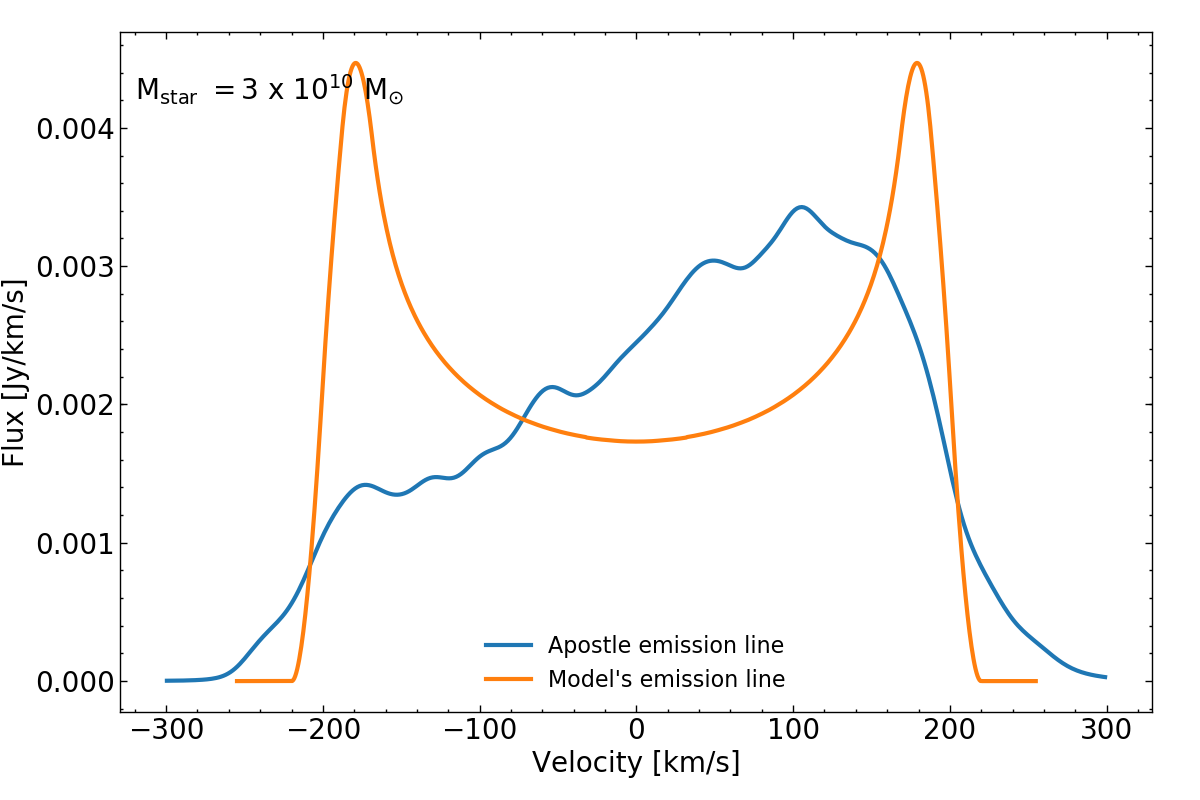}
        \vspace{0.5cm}
        \includegraphics[scale = 0.3]{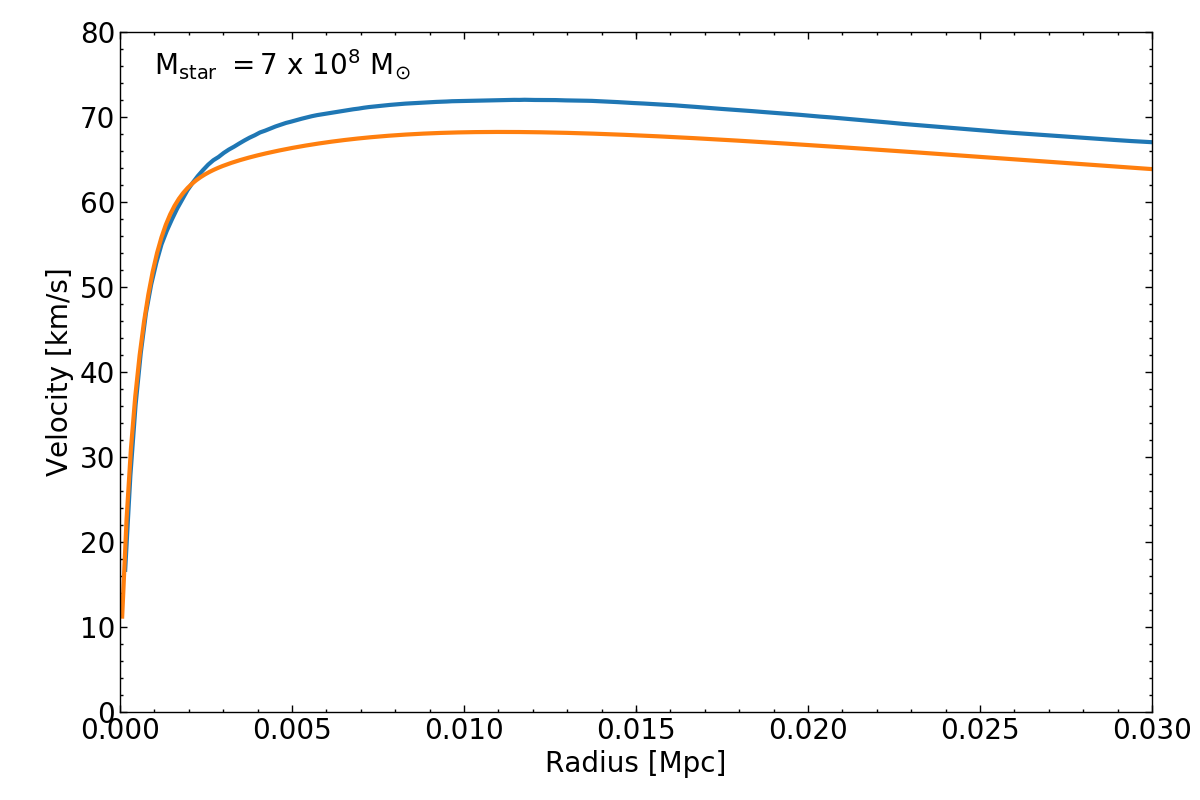}
        \hfill
        \includegraphics[scale = 0.3]{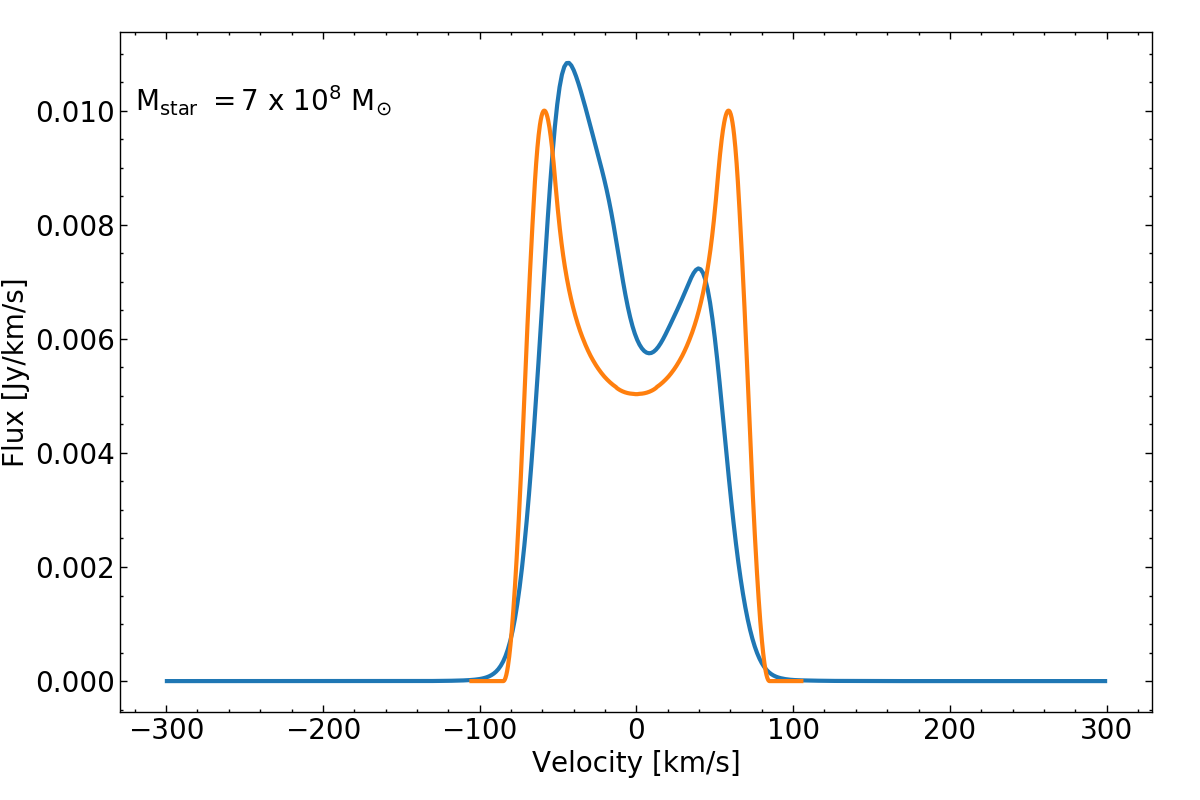}
        \vspace{0.5cm}
        \includegraphics[scale = 0.3]{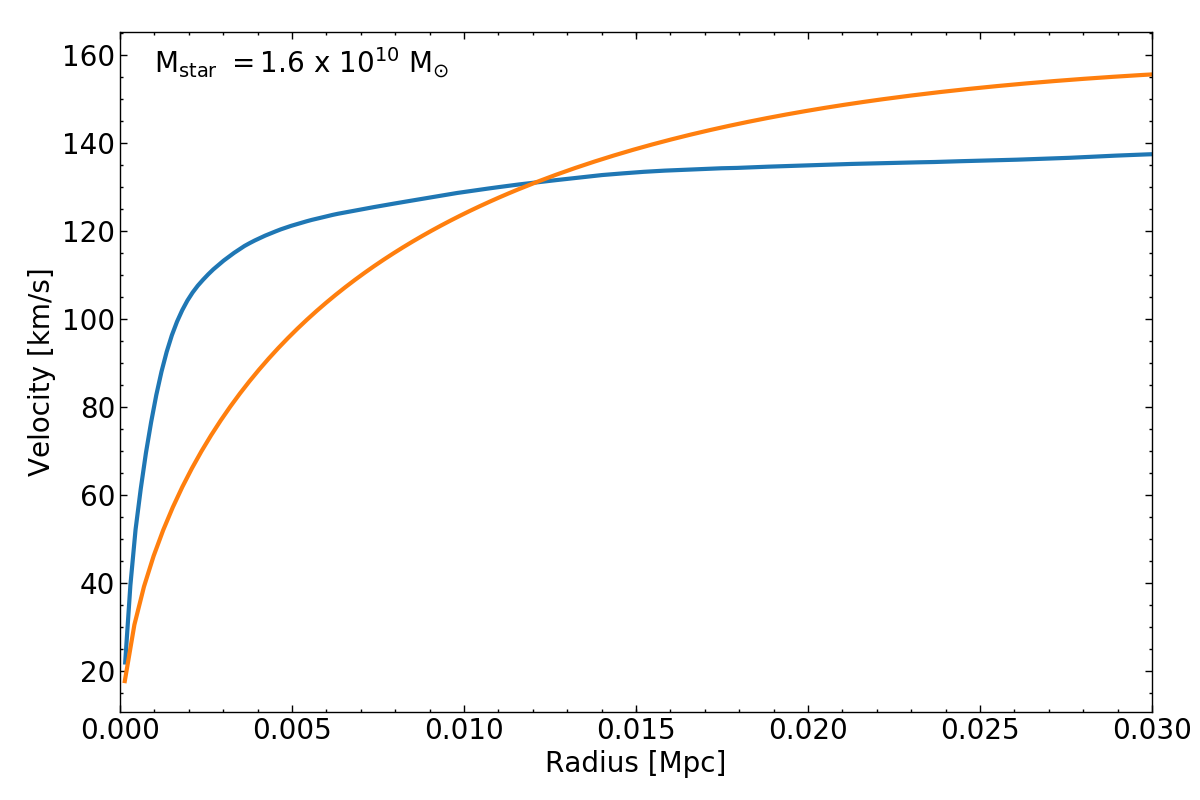}
        \hfill
        \includegraphics[scale = 0.3]{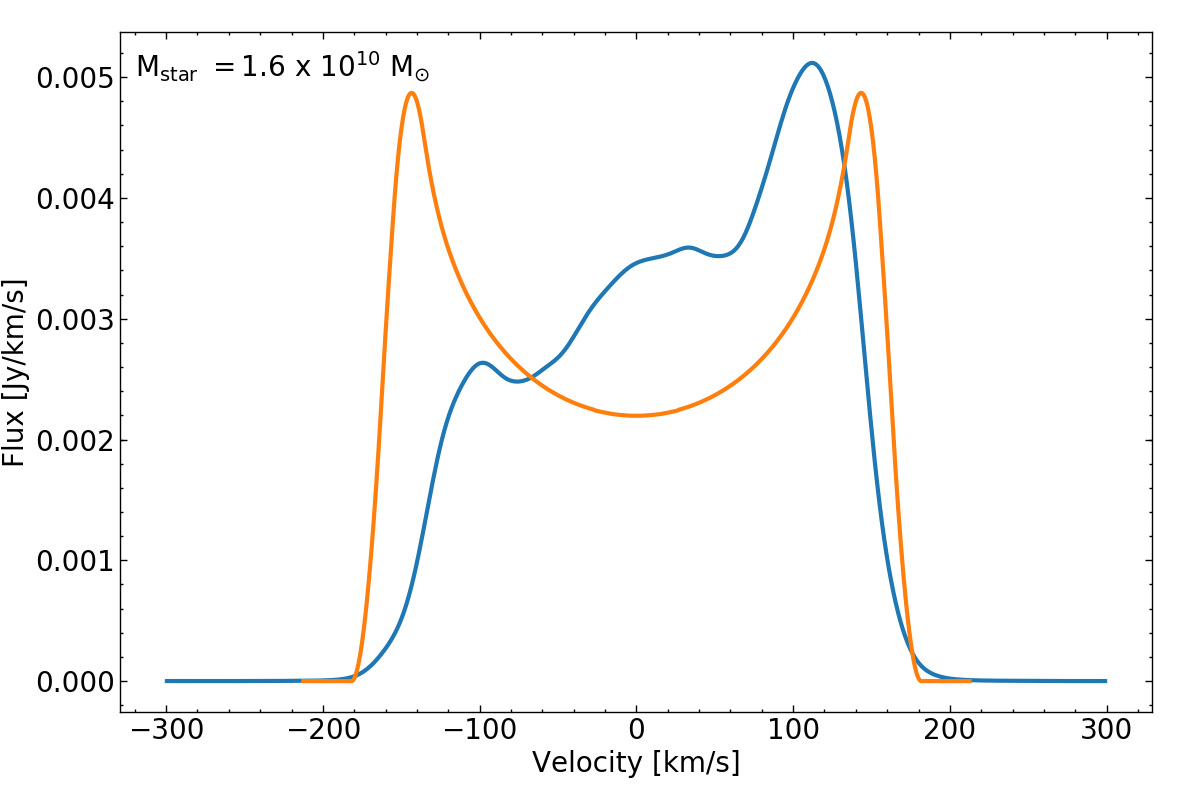}
  \end{minipage}
\caption{Rotational Velocity curves (left panel) and corresponding \hi\ emission line profiles (right panel) from the APOSTLE simulations compared to our model. The blue and orange lines correspond to Apostle and our model results, respectively, with the stellar mass of the galaxies as labelled. We show three examples of a galaxy in which our model does poorly (top panel), does well (middle) and an intermediate case (bottom panel).}
\label{fig:apostle-rotation-curves}

\end{figure*}


\begin{figure*}
\minipage{0.5\linewidth}
  \includegraphics[width=\linewidth]{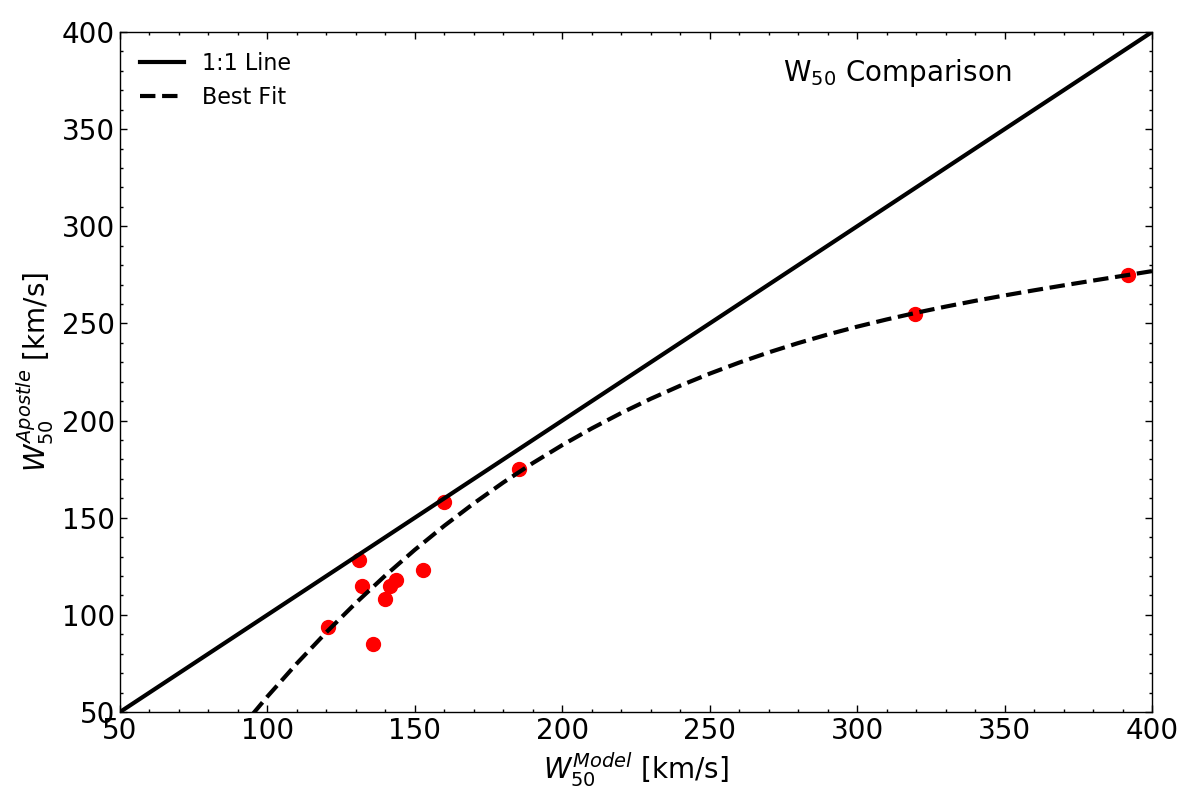}
\endminipage\hfill
\minipage{0.5\linewidth}
  \includegraphics[width=\linewidth]{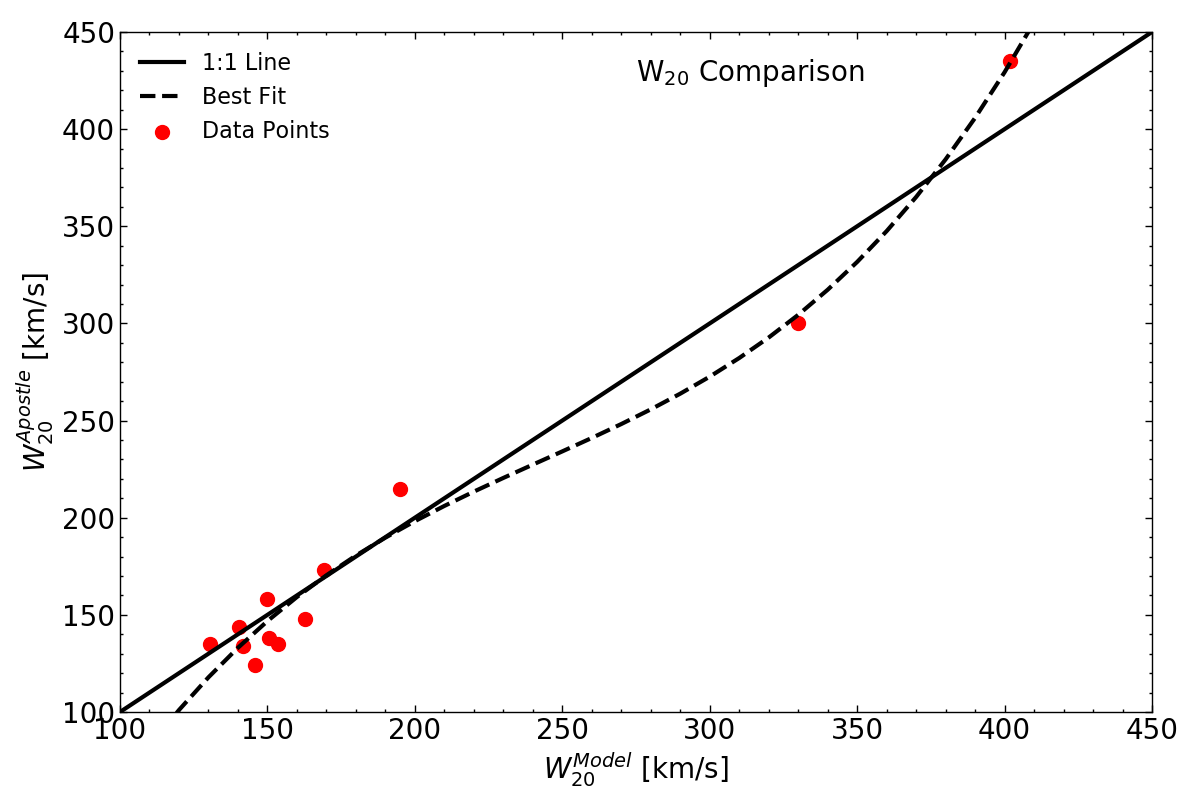}
\endminipage

\caption{Comparison of the \wtwenty\ and \wfifty\ measurements taken for the \hi\ emission lines in APOSTLE and that produced by our idealised model, with the points being individual galaxies, the solid line being the 1:1 ratio and dashed line being the best spline fit. It should be noted that \wtwenty\ measurements agree better between the hydrodynamical simulations and our idealised model than for \wfifty. This is because most of the \hi\ emission line spectra in APOSTLE are asymmetric, which affects \wfifty  more than \wtwenty.}
\label{fig:apostle-w20-w50}
\end{figure*}


The APOSTLE cosmological hydrodynamical simulations  \citep{Sawala2016ThePuzzles} are a suite of twelve `zoom-in' volumes evolved with the code and models developed and calibrated for the EAGLE project \citep{2015MNRAS.446..521S,2015MNRAS.450.1937C}. The volumes are selected to resemble the Local Group of galaxies in terms of the masses of two central objects -- analogous to the Milky Way and M~31, their separation, relative velocity, and relative isolation from other massive systems. Each volume is evolved at 3 resolution levels. The lowest level L3 is similar to the fiducial EAGLE resolution \citep[e.g. L0025N0376 in the nomenclature of][]{2015MNRAS.446..521S}, with a gas particle resolution of $\sim 10^6\,{\rm M}_\odot$ and gravitational softening of $\sim 700\,{\rm pc}$. The two higher resolution levels each decrease the particle resolution by a factor of $\sim 8$, for a gas particle mass at maximum resolution L1 of $\sim 10^4\,{\rm M}_\odot$, and a gravitational softening of $\sim 130\,{\rm pc}$. The code uses the ANARCHY implementation \citet{2015MNRAS.454.2277S} of pressure-entropy smoothed particle hydrodynamics \citep{2013MNRAS.428.2840H}, and includes prescriptions for radiative cooling \citep{2009MNRAS.393...99W}, an ionizing background \citep{2001cghr.confE..64H}, star formation \citep{2004ApJ...609..667S,2008MNRAS.383.1210S}, supernovae and stellar mass loss \citep{2009MNRAS.399..574W}, energetic feedback from star formation \citep{2012MNRAS.426..140D} and AGN \citep{2009MNRAS.398...53B,2015MNRAS.454.1038R}. Full details of the model and calibration are available in \citet{2015MNRAS.446..521S,2015MNRAS.450.1937C}, and of the APOSTLE simulations in \citet{Sawala2016ThePuzzles,2016MNRAS.457..844F}. APOSTLE uses the REFERENCE calibration of the EAGLE model \citep[see][]{2015MNRAS.446..521S}, and the WMAP7 cosmological parameters \citep{2011ApJS..192...18K}.

The code {\sc martini} \footnote{https://github.com/kyleaoman/martini} was used to produce neutral hydrogen (H\,{\sc i}) emission line profiles for a selection of galaxies from the APOSTLE simulations. A detailed description of an earlier version is available in \citet{Oman2019}. The hydrogen ionization fraction of each simulation particle is estimated following \citet{2013MNRAS.430.2427R}; the neutral hydrogen is further partitioned into atomic and molecular gas following \citet{Blitz2006}. Each particle contributes flux to the spectrum distributed as a Gaussian centered at the particle velocity, with a width specified by $\sqrt{k_B T / m_p}$, where $k_B$ is Boltzmann's constant, $T$ is the particle temperature, and $m_p$ is the particle mass, and an amplitude proportional to the neutral hydrogen mass of the particle. The galaxies are placed edge-on ($i=90^\circ$) at a fiducial distance of $D=10\,{\rm Mpc}$, with a systemic velocity of $H_0 D$, with $H_0=70\,{\rm km}\,{\rm s}^{-1}\,{\rm Mpc}^{-1}$. The galaxies are selected morphologically to host gas discs, and to span a range in total (dynamical) mass, with $14$ between $60<V_{\rm max}/{\rm km}\,{\rm s}^{-1}<120$ and $2$ with $V_{\rm max}\sim 200\,{\rm km}\,{\rm s}^{-1}$, where $V_{\rm max}$ is the maximum of the circular velocity curve. Other quantities required as inputs for our model were measured directly from the simulation particle properties -- specifically, virial mass of the halo,\hi\ and stellar mass of galaxy and half-mass stellar and gas radii for the galaxy.  


We build \hi\ emission lines following the procedure described in $\S$~\ref{sec:Modelling} using the input global properties specified above. On the other hand, the \hi\ emission lines from APOSTLE make full use of the complex geometry and non-circular motions that are predicted by the simulation. We compare our idealised model with the \hi\ emission lines predicted by APOSTLE with the aim of understanding the systematic effects introduced by our assumptions with respect to more realistic \hi\ line profiles. We used $13$ dwarf galaxies and $2$ Milky way sized galaxies to compare our models.

In Figure~\ref{fig:apostle-rotation-curves}, we compare the \hi\ emission for 3 example galaxies, highlighting cases in which our idealised model provided a poor and a good representation of the \hi\ emission line (top and middle panels, respectively), with the bottom panel showing an intermediate case.

We find that for some galaxies the estimates of our model and the \hi\ generated by the simulation show comparable widths and rotation curves but for others our model produces a rotation curve that flattens are smaller radii. When we compare the \wfifty\ and \wtwenty\ (see Figure~\ref{fig:apostle-w20-w50}), we notice that for galaxies with a higher mass or higher velocity and symmetric double-horned profile shape, we produce measurements that are close to the APOSTLE ones. We find better agreement in our \wtwenty\ values than the \wfifty\ estimates. The cause for this is the asymmetry of the lines in the APOSTLE simulated galaxies, which leads to systematically different \wfifty\ estimates (due to the heights of the lines), which play a lesser role on \wtwenty. This suggests that \wtwenty\ should be a more stable, reliable estimate of the dynamical mass, in agreement with the inferences of \citet{McGaugh2011}.

The reason why the \hi\ emission lines in APOSTLE are so asymmetric and whether that agrees with observations is unclear. \citet{Oman2019} studied the velocity profiles of APOSTLE dwarf galaxies, finding significant contribution from non-circular motions in addition to the purely circular velocity. \citet{Sales2017} found that APOSTLE dwarf galaxies may be significantly deviating from the measured Tully-Fisher relation of \citet{Papastergis2016}. The latter may be an indication that feedback effects are too strong in APOSTLE. However, further research on the \hi\ line profiles of APOSTLE galaxies is required before we can make draw robust conclusion.

Equations~\ref{eq:modified-w50} and \ref{eq:modified-w20} show spline fits to the relations shown in Figure~\ref{fig:apostle-w20-w50}. These equations could be used as an approximation to the deviations of \wfifty\ and \wtwenty\ from our idealised model.

\begin{equation}
\begin{multlined}
    W_{\rm 50_{\rm Apostle}} = 6.02\times10^{-6}\times W_{\rm 50_{\rm Model}}^3 -7.04\times10^{-3}\times W_{\rm 50_{\rm Model}}^2 \\ + 2.98\times W_{\rm 50_{\rm Model}} - 176.04
\end{multlined}    
    \label{eq:modified-w50}
\end{equation}

\begin{equation}
\begin{multlined}
    W_{\rm 20_{\rm Apostle}} = 2.40\times10^{-5}\times W_{\rm 20_{\rm Model}}^3 -1.75\times10^{-2}\times W_{\rm 20_{\rm Model}}^2 \\ - 4.93\times W_{\rm 20_{\rm Model}} -280.69
\end{multlined}    
    \label{eq:modified-w20}
\end{equation}



\bsp	
\label{lastpage}
\end{document}